\documentclass[a4paper,12pt]{article}

\usepackage[usenames,dvipsnames,svgnames,table,x11names]{xcolor}
\usepackage[export]{adjustbox}

\usepackage[percent]{overpic}
\usepackage{kpfonts}
\usepackage{bbm}
\usepackage{geometry}
\usepackage{anyfontsize}
\usepackage{t1enc}
\usepackage{feynmp-auto}
\usepackage{hyperref}
\usepackage{braket}
\usepackage{packages-edited}
\usepackage{definitions}
\usepackage{simeontex}
\usepackage{tikz}
\def\bbq{{\bf Q}}
\hypersetup{
     colorlinks   = true,
     citecolor    = magenta
}
\usepackage{cite}

\def\dagsqd{^{\dagger 2}}

\def\ddsq{\dagsqd}
\def\ddsqd{\dagsqd}

\def\app#1#2{%
  \mathrel{%
    \setbox0=\hbox{$#1\sim$}%
    \setbox2=\hbox{%
      \rlap{\hbox{$#1\propto$}}%
      \lower1.1\ht0\box0%
    }%
    \raise0.25\ht2\box2%
  }%
}

\renewcommand{\ll}{_}
\def\muchlessthan{< \hskip-.05in <}
\renewcommand{\gg}{\nabla}
\def\muchgreaterthan{> \hskip-.05in >}
\def\dket#1{\big |\cc {#1} \cc\cc \big \rangle \hskip-.03in  \big \rangle}
\def\dbra#1{\big \langle \hskip-.03in  \big\langle \cc\cc  {#1} \cc \big |}
\def\nbdbra#1{\big \langle \hskip-.03in \big\langle \cc\cc  {#1}}

\newcommand{\uset}[2]{\underset{\textcolor{magenta}{\substack{#2}}}{#1}}
\newcommand{\puu}{\partial_{\uparrow\uparrow}}

\renewcommand{\IR}{{\mathbbm{R}}}

\def\D{\Delta}

\makeatletter
\g@addto@macro\bfseries{\boldmath}
\makeatother

\def\bdag{b\dag}
\def\adag{a\dag}

\def\ddp{{\tilde{\Delta}}}
\def\qqq{{\bf q}}

\newenvironment{PurpleEnv}%
{\color{Purple}}%
{\color{Black}}

{\color{Blue}}%
{\color{Black}}

{\color{Purple}}%
{\color{Black}}
{\color{Red}}%
{\color{Black}}

\def\bbp{\begin{PurpleEnv}}
\def\eep{\end{PurpleEnv}}

\begin{document}

\setcounter{tocdepth}{2}

\begin{titlepage}
\begin{flushright}
IPMU-17-0015\\
CALT-TH-2017-032\\
\end{flushright}
\vspace{8 mm}
\begin{center}
  {\large \bf Operator Dimensions From Moduli
  }
\end{center}
\vspace{2 mm}
\begin{center}
{Simeon Hellerman$^1$, Shunsuke Maeda$^{1,2}$ and  Masataka Watanabe$^{1,2}$}\\
\vspace{6mm}
{\it $^1$Kavli Institute for the Physics and Mathematics of the Universe (WPI),\\
 University of Tokyo,\\
 Kashiwa, Chiba  277-8582, Japan\\}
 \vspace{6mm}
{\it $^2$Department of Physics, Faculty of Science,\\
University of Tokyo, Bunkyo-ku, Tokyo 133-0022, Japan\\}
 \vspace{6mm}

\end{center}
\vspace{-10 mm}
\begin{center}
{\large Abstract}
\end{center}
\noindent
 We consider the operator spectrum of a three-dimensional ${\cal N} = 2$ superconformal field theory with a moduli space of one complex dimension, such as the fixed point theory with three chiral superfields $X,Y,Z$ and a superpotential $W = XYZ$.  By using the existence of
an effective theory on each branch of moduli space, we calculate the anomalous dimensions of certain low-lying operators carrying 
large $R$-charge $J$.  While the lowest primary operator is a BPS scalar primary, the second-lowest scalar primary  is in a semi-short 
representation, with dimension exactly $J+1$, a fact that cannot be seen directly from the $XYZ$ Lagrangian.  The third-lowest
scalar primary lies in a long multiplet with dimension $J+2 - c_{-3} \cc J\uu {-3} + O(J\uu{-4})$, where $c_{-3}$ is an unknown
positive coefficient.  The coefficient $c_{-3}$ is proportional to the leading superconformal
interaction term in the effective theory on moduli space.  The positivity of $c_{-3}$ does \textit{not} follow from supersymmetry, but rather
from unitarity of moduli scattering and the absence of superluminal signal propagation in the effective dynamics of the complex modulus. 
We also prove a general lemma, that scalar semi-short representations form a module over the chiral ring in a natural way, by ordinary
multiplication of local operators.  Combined
with the existence of scalar semi-short states at large $J$, this proves the existence of scalar semi-short states at \rwa{all}
values of $J$.  Thus the combination of ${\cal N}=2$ superconformal symmetry with the large-$J$ expansion is more powerful than the sum of its parts.

\vspace{1cm}
\begin{flushleft}
\today
\end{flushleft}
\end{titlepage}
\hypersetup{linkcolor=black}

\tableofcontents
\hypersetup{linkcolor=PaleGreen4}
\newpage

\section{Introduction} 

Generic CFT need not be weakly coupled, in the sense of lying in families -- discrete or continuous -- with limits that can be reduced
to weakly interacting free fields or solved exactly in some more general way. However even
individual theories that are strongly coupled, may have \rwa{families of observables} with limits that are weakly coupled in this
sense.  The simplest such limit to study is that of large global charge.

 In \cite{Hellerman:2015nra} and related work \cite{Alvarez-Gaume:2016vff, Monin:2016jmo, Loukas:2016ckj, Hellerman:2017efx}
 it was noted that three dimensional CFT with global symmetries simplify in this sense in 
some familiar examples, when one considers the dimensions of low-lying operators of large global charge $J$.  The theories
considered in \cite{Hellerman:2015nra} are the critical $O(2)$ model and the infrared fixed point
of the ${\cal N} = 2$ supersymmetric theory with a single chiral superfield $\Phi$ and superpotential $W = \Phi\uu 3$.
In each case, the operator dimension of the lowest operator with charge $J$ is a scalar operator with dimension
$\Delta_J = c_{{3\over 2}} J\uu {+{3\over 2}} +  c_{{1\over 2}} J\uu {+{1\over 2}} -0.0093 + O(J\uu{-\hh})$\, where
the unknown coefficients $c_{{3\over 2}}, c_{{1\over 2}}$ may differ between theories and the $J$-independent term is universal.\footnote{The value was corrected
\cite{Monin:2016bwf} from the one originally appearing in \cite{Hellerman:2015nra}, which suffered from a misuse of $\zeta$-function regularization.}   The common form of
the large-$J$ expansions for $\Delta_J$ follows from the fact that the two theories are both described by 
an effective theory in the same universality class.  The effective theory describes a single compact scalar $\chi \simeq \chi + 2\pi$ transforming
as $\chi \to \chi + ({\rm const.})$ under the global symmetry.  This effective theory can be used to calculate dimensions of operators 
using radial quantization on the sphere of radius $r$, taking the Wilsonian cutoff $\Lambda$ to satisfy $r\uu{-1} \muchlessthan 
\Lambda \muchlessthan {{\sqrt{J}}\over r}$.  Then both quantum effects and higher-derivative terms are suppressed by powers
of $J$, and the renormalization group equation at the conformal fixed point forces all $\Lambda$ dependence cancel in 
observables, order by order in $J$.  Crucially, conformal invariance is a symmetry of the effective theory at the quantum level,
which constrains the $\Lambda$-independent terms to be classically scale invariant (indeed, Weyl invariant) and further determines
the $\L$-dependent terms uniquely in terms of the $\L$-independent terms.

The $J\uu{+{3\over 2}}$ scaling in the $O(2)$ and $W = \Phi\uu 3$ models is a consequence of the fact that neither theory has a continuous family of Lorentz-invariant vacua
on flat spatial slices. 
As a result, the curvature of the space is irrelevant for high-energy states on the sphere, and so the relationship between the energy density ${\cal H}\equiv T_{00}$ and the charge density $\r \equiv J_0$ in the large-$J$ ground state can only be ${\cal H} \sim |\r|\uu{+{3\over 2}}$.  

In the case of the ${\cal N} = 2$ superconformal $W = \Phi\uu 3$ theory, the $J\uu{+{3\over 2}}$ scaling encodes the fact that SUSY is strongly spontaneously broken at large
$R$-charge.  That SUSY must be broken follows from the fact that the chiral ring truncates; the only BPS scalar primaries are $ 1$
and $\phi$ itself.  However the fact that the breaking is parametrically large at high $J$, cannot be understood in terms of
the structure of the chiral ring, and only the use of the effective theory can uncover it.

The case of an infinite chiral ring is different.  For ${\cal N} = 2$ SCFT with infinite (finitely generated) chiral ring, there is 
a moduli space of Lorentz-invariant vacua on $\IR\uu 3$, whose holomorphic coordinate ring is the "radical" of the chiral ring --
the chiral ring modulo its nilpotent elements     \cite{Luty:1995sd}.  Conversely, if there is a $k$-complex-dimensional moduli space, then 
it is described by a nilpotent-free chiral ring with $k$ algebraically independent generators, or $k+\ell$ generators with
$\ell$ relations.  

The moduli space of vacua implies a degenerate spectrum when the
curvature vanishes, and consequently the curvature is always relevant in the relationship between ${\cal H}$ and $\r$ on the sphere.  For a theory with a moduli space of vacua, low-lying states of large $R$-charge satisfy
a relationship of the form ${\cal H} \sim\big ( {{{\tt Ric}}\over 2} \big ) \uu{+{1\over 2}} \cc |\r|$, 
where ${\tt Ric} $ is the Ricci scalar curvature of the spatial slice.  In radial quantization, this translates into a relationship
$\Delta_J \sim  +1 \cdot |J|\uu 1$, a relationship that is exact for the lowest operator with charge $J$, which is always BPS if a moduli space exists
and $J$ satisfies an appropriate quantization condition.

Possibly the simplest interacting theory with a moduli space is the $XYZ$ model, the ${\cal N} = 2$ superconformal
infrared fixed point of three free chiral superfields $X,Y,Z$ 
perturbed by a superpotential $W = g \cc XYZ$ where $g$ is a coupling constant with dimension $[g] = [{\rm mass}]\uu{+\hh}$.
At scales $E \muchlessthan g\sqd$ this theory flows to a superconformal theory that is strongly coupled: The the anomalous dimensions and OPE coefficients of the fields\footnote{We will use the same
notation for the superfields $X,Y,Z$ and their $\theta =\bar\theta= 0$ bosonic components.} $X,Y,Z$ are all of $O(1)$.  The theory has exactly one marginal operator that
breaks the $U(1)\uu 3$ symmetry and lowers the dimension of the moduli space (as shown by a Leigh--Strassler type argument  \cite{Strassler:1998iz}),
but no marginal operators neutral under the full global symmetry.
The chiral
ring of this theory is one-dimensional, corresponding to the case where $k=1$ and $\ell = 2$.  The three generators $X,Y,Z$
obey the relation $XY = XZ = YZ = 0$, so that the moduli space consists of three branches, freely generated by $X, Y$,
and $Z$ respectively.  

\begin{sloppypar}
That is, the chiral ring consists of linear combinations of the elements, $ \{X\uu p, Y\uu p, Z\uu p, ~\forall  p \geq 0\}$.
This theory has three independent $U(1)$ global symmetries $U(1)_{X,Y,Z}$ under which $X,Y,$ and $Z$ carry charge $+1$, respectively,
and the $R$-charge $J_R$ is a linear combination of the three, 
\begin{align}
\begin{split}
J_R \equiv {2\over 3}(J_X + J_Y + J_Z)\ .
\end{split}
\end{align}
Scalar superconformal primary operators are in the chiral ring if they satisfy the BPS bound $\Delta = J_R$.  If an
operator is not in the chiral ring, but satisfies $(\Delta - J_R) / J_R \muchlessthan 1$, we ought to be able to think of
them as "near-BPS", and their anomalous dimensions should be computable in large-$J$ perturbation theory.
\end{sloppypar}

The first examples of such a perturbation theory known to the authors are    \cite{Berenstein:2002jq,Gross:2002su}.  In those
papers the authors used the additional simplifications of ${\cal N} = 4$ SUSY in four dimensions, as well as the planar
approximation at large $N$.  In this paper we will show that near-BPS operator dimensions can be calculated straightforwardly
in a large-$J$ expansion in the $XYZ$ model, which lacks these additional simplifications.

In particular, we will extend the approach of     \cite{Hellerman:2015nra} to study the anomalous dimensions of near-BPS scalar operators 
with large $R$-charge on the "$X$ branch", that is, operators with $J_R ,~ J_X \muchgreaterthan 1$, and
$J_X - {3\over 2} J_R$ and $\Delta - J_R$ of order 1.  We will quantize the theory on $S\uu 2$ spatial slices and
use the effective field theory of the $X$ branch of moduli space, to compute the operator dimension realized as the energy of
the state on the sphere, \it via \rm the state-operator correspondence.  At large $J$, the moduli space effective theory
becomes a controlled tool: Both higher-derivative operators and Feynman diagrams with loops have their effects suppressed by
powers of $J$.

We find some interesting results: 

\bi
\item{After the lowest operator, which is BPS, the next-lowest operator with the same charge is also a scalar and lies in a "semi-short" multiplet  -- its $\Qbar\sqd$ descendant is absent and its dimension is precisely $J+1$.}
\item{The third-lowest operator with the same charges lies in a long multiplet and receives corrections to its dimension 
of order $J\uu{-3}$.}
\item{The $J\uu{-3}$ correction comes from a single insertion of the lowest-derivative super-Weyl-invariant interaction term in the effective theory on moduli space.}
\item{The coefficient of this effective term is not perturbatively calculable, but its sign is positive definite by virtue of the causality
constraint discussed in     \cite{Adams:2006sv}.  As a result,
the $J\uu{-3}$ correction to the energy of the third-lowest scalar primary
operator is negative definite. 
}
\ei

\section{Effective theory of the $X$ branch}\label{XBranchEFT}

As in the case of the $O(2)$ model in \cite{Hellerman:2015nra}, we begin by observing that the Wilsonian action at large values of
the fields $X,Y,Z$ has an expansion in powers of the cutoff over the UV scale defined by the scalar vevs themselves.  In
this regime the loop contributions to the RG equation, both the finite and $\L$-dependent parts,
are parametrically smaller than the action of a classical scale transformation,
and the RG fixed point equation becomes the condition for classical scale invariance, with calculable corrections that simply
determine the $\L$-dependent terms from the $\L$-independent terms.  For the particular directions where one of the three
fields is nonzero and the other two vanish, the other two fields are massive, with masses above the cutoff, and one can obtain
an effective action for one of these fields alone.

\heading{Structure of the effective action}

The effective theory of the $X$ modulus has a relatively simple structure.  In flat space, terms can be understood as
full-superspace integrals,
\begin{align}
\begin{split}
\Delta {\cal L} = \int \cc d\sqd \th d\sqd \thb \cc {\cal I}\ ,
\end{split}
\end{align}
where ${\cal I}$ is an operator with $J_X = J_R = 0$.  

Terms in the effective Lagrangian can be classified into two types: classical and quantum terms.  Classical terms are independent of the
cutoff $\Lambda$ and invariant under the Weyl transformation $X \to \operatorname{exp}\left(\frac{2\sigma}{3}\right) X$, with the
superpartner of $X$ transforming as $\psi^X\to\operatorname{exp}\left(\frac{7\sigma}{6}\right)\psi^X$.
The scaling dimension of ${\cal I}\ll{\rm classical}$ must be exactly $1$.

Quantum terms are entirely dependent on the form of the regulator and scale as positive powers of the cutoff scale $\L$.  They are
not Weyl-invariant or even scale invariant as terms in the action; scale invariance is explicitly broken by the $\L$-dependence.
These terms are of the form
\begin{align}
\begin{split}
{\cal I} = \L\uu q \cc {\cal I}_{1-q}\ ,
\end{split}
\end{align}
where ${\cal I}_{1-q}$ is an operator of dimension $1-q$ and $q > 0$.

It is important to note that the condition $q > 0$ satisfied by the quantum terms
is not a universal rule in effective field theories: When we integrate out a shell of modes,
between $\L$ and $\L - \d \L$, the propagators are ${1\over{\L\sqd}}$ and naively it would seem that $\L$ can appear
to negative powers in the Wilsonian action.  Rather, the $q > 0$ rule follows from the fact that the effective theory on moduli space is infrared-free.  The 1PI effective action for $X$, expanded around
a nonzero vev, is therefore convergent, since $X$ is an observable in the infrared theory.  But the 1PI effective action
is nothing more than the $\L \to 0$ limit of the Wilsonian effective action.  So the Wilsonian action must be
finite in the $\L \to 0$ limit.

There is less to the quantum terms than meets the eye.  
We are treating our theory as a Wilsonian theory in a perturbatively controlled regime by taking $E_{{\rm IR}} \muchlessthan
\L \muchlessthan E_{{\rm UV}}$, with $E_{{\rm UV}} = |X|\uu{+ {3\over 2}}$.  The condition of conformal invariance dictates
that the effect of integrating out a shell of modes and lowering the cutoff from $\L$ to $\L - \d \L$ must be exactly cancelled
by rescaling the momenta by a factor of $(1 - {{\d \L}\over{\L}})\uu{-1}$, with a redefinition of the fields to restore the normalization
of the kinetic term.  This allows us to write a renormalization group equation which can be solved to derive
the coefficients of the $\L$-dependent quantum terms from the $\L$-independent classical action, order by order in 
${{\L}\over{E_{{\rm UV}}}} = {{\L}\over{|X|\uu{3\over 2}}}$.  The quantum terms, then, comprise a sort of epiphenomenon: Once
the regulator has been fixed and the conformal invariance of the underlying theory is taken as an input, the cutoff-dependent
terms contain no independent information.

Concretely, the RG evolution of the classical action is of the form\footnote{There can also in principle
be dependences of the form $\L\uu{q_i} \cc [{\tt ln}(\L / E_{{\rm UV}})]\uu {s_i}$ with $q\ll i > 0$.  These can be incorporated into the RG
equation but for simplicity we omit them.}
\begin{align}
\begin{split}
\d\ups{{\rm RG}} {\cal L}\uprm{{\rm classical}} = \sum_i \L\uu{q_i} \cc {\cal O}_i\ ,
\end{split}
\end{align}
where ${\cal O}_i$ is an operator of dimension $\Delta_i \equiv 3 - q_i$.  Then the fixed point equation for RG evolution
allows us to solve for ${\cal L}\upp{a+1}$ in terms of ${\cal L}\upp a$ by
\begin{align}
\begin{split}
\L{{\d\uprm{RG}}\over{\d \L}} {\cal L}\upp a = \sum\ll{\D\ll c \neq 3} \cc (3 - \D\ll c) \cc  \L\uu{3 - \D\ll c} \cc {\cal L}\upp{a+1}\ll{\D\ll c}\ ,
\end{split}
\end{align}
where ${\cal L}\upp{a+1}\ll{\D\ll c}$ is the set of terms in the Lagrangian with canonical scaling dimension $\D\ll c$.

We do not need to know the concrete form of the quantum terms at all for most practical purposes.  In
correlation functions, their only
role is to cancel the $\L$-dependence from quantum amplitudes order by order in $E_{{\rm IR}} / |X|\uu{3\over 2}$.  In
practice, we can simply quantize the classical action with a (sufficiently supersymmetric) cutoff,
and add local counterterms with positive powers of $\L$ to cancel any divergences.  Since the underlying
theory is conformal, there is no danger of getting the wrong answer by doing this.

For purposes of tree-level amplitudes we need not consider the $\L$-dependent terms at all, and for one-loop
amplitudes it is simplest to use a scale-free regulator such as $\zeta$-function or dimensional regularization.  We
therefore need not consider the $\L$-dependent terms further in this paper.

\heading{Weyl invariance and super-Weyl invariance}

Weyl invariance constrains terms more strongly than simple scale invariance, and super-Weyl invariance constrains them more strongly
still.  Since the original CFT is Weyl invariant and 
can be formulated on an arbitrary geometry\footnote{at least, a smooth one of nonnegative scalar curvature},
the same must be true of the effective theory of the $X$ branch.  This turns out to be rather constraining for possible
effective operators.

First, let us see why certain low-derivative terms that are scale invariant cannot be given a Weyl-invariant completion and therefore
cannot appear as terms in the effective action of a CFT, even without considering constraints due to SUSY.  

Defining the field \footnote{This transformation is innocuous when calculating energies, but potentially subtle when computing two-point functions of large-charge operators, when the classical solution can attain $0$ and $\infty$ and
the physics may be sensitive to the singular branch points there.}
\bbb
\phi \equiv X\uu{3\over 4}\ ,
\een{PhiFieldDef}
we see that the leading scale-invariant term in three dimensions would
be $|\phi|\uu 6 = |X|\uu \frac{9}{2}$. 
This term is also Weyl-invariant although it is of course disallowed by supersymmetry.  Note
that when we say a term $\co$ in the Lagrangian is scale-invariant or Weyl-invariant, we mean this as a shorthand that
the term transforms as a tensor of weight $3$, so that $\int \cc \sqrt{|g|} \cc d\uu 3 x \cc \co$ is truly Weyl-invariant.

The next term would be
the kinetic term $(\nabla_\m \phi)(\nabla\uu\m \phb)$.  This term is scale-invariant but \rwa{not} Weyl-invariant and therefore
not conformally invariant either.  However it has a Weyl-invariant completion
obtained by adding the conformal coupling 
to the Ricci scalar, $+ {1\over 8} {\tt Ric}_3 \cc |\phi|\sqd$.

\heading{Leading interaction term}
Now let us discuss the interaction term and its effects on the spectrum of the $X$ branch.

There are no Weyl-invariant bosonic operators with three derivatives. This is immediately clear on the basis of Lorentz invariance and parity.
At the four-derivative level, there is a unique Weyl-invariant operator with four-derivatives    \cite{Paneitz2008,MR1190438,MR1190439,Gover:2002ay,Gover:2003am} which in
flat space takes the form of a supersymmetrized operator of Fradkin-Tesytlin-Paneitz-Riegert (FTPR) type \cite{Fradkin:1981jc,Fradkin:1982xc,Paneitz2008,Riegert:1984kt}:
\bbb
\co_{\rm FTPR} \equiv
\frac{1}{\bar\phi}
\partial^2 \partial^2 \frac{1}{\phi}
 , 
\een{FTPRFlatSpace}
The curvature couplings of the FTPR operator have been worked out     \cite{Paneitz2008}, and take the form
\begin{align}\begin{split}\label{PaineitzCurvatureCouplings}
&\co_{\rm FTPR}
\\& 
\equiv
\frac{1}{\bar\phi}\left[
\nabla^2\nabla^2
+\nabla_\mu\left(\frac{5}{4}g^{\mu\nu}R-4R^{\mu\nu} \right)\nabla_\nu
-\frac18\left(\nabla^2 R\right)+R^{\mu\nu}R_{\mu\nu} -\frac{23}{64}R^2
\right]\frac{1}{\phi}
,\end{split}
\end{align}
where $R_{\mu\nu}=\left(\mathtt{Ric}_{3}\right)_{\mu\nu}$ is the
Ricci tensor and $R=\mathtt{Ric}_{3}$ is the Ricci scalar.

For purposes of this paper, we need only know the Lagrangian on $S\uu 2 \times \IR$, which is conformally flat.  So we could have
worked out the curvature couplings for this particular geometry by Weyl-transforming the flat-space FTPR term under
the transformation
\begin{align}
\begin{split}
ds\sqd_{{S\uu 2 \times \IR}} = {{r\sqd}\over{|w|\sqd}} \cc ds\sqd_{{\IR\uu 3}}\ ,
\end{split}
\end{align}
where $w\uu\m$ are the linear coordinates on $\IR\uu 3$ and $r$ is the radius of the sphere.
The curvature couplings derived this way agree with the general form in \rr{PaineitzCurvatureCouplings}
In principle, there may be terms in the effective theory that are Weyl-invariant but do not have a super-Weyl-invariant completion.
However the super-Weyl-invariant completion of the FTPR operator exists and has recently been written down    \cite{Kuzenko:2015jda}. 
 In flat space, it can be written as the superspace invariant
\begin{gather}
\begin{array}{c}\displaystyle
\label{SuperFTPRFlatSpace}
\co_\textrm{{super-FTPR}} \equiv \int \cc d\uu 2 \th d^2 \bar\theta \cc {\cal I}_{\textrm{super-FTPR}}\ 
,\qquad
{\cal I}_\textrm{ super-FTPR} \equiv 
 \frac{\partial_\mu\Phi \partial^\mu\bar{\Phi}}{\left(\Phi\bar{\Phi}\right)^2},
\end{array}
\end{gather}
where $\Phi\equiv \phi + \sqrt{2}\theta\psi+\cdots$ is a chiral superfield,
whose complete expression is given in (\ref{ChiralSFExpansion}). 

By coupling this action to a background supergravity multiplet (see for instance     \cite{Festuccia:2011ws,Dumitrescu:2012ha}), we should in principle be able to derive the
general curvature couplings of the fermions as well.  However in practice, working out the component form of the super-FTPR
term from the curved superspace expression is quite cumbersome.  Since we only interested for purposes of this paper in 
the case of $S\uu 2 \times \IR$, it is more efficient to Weyl-transform the action directly from the flat-space expression.  For a 
sphere of radius $r$, we obtain:
\begin{align}
\mathcal{O}_{\text{super-FTPR}}&=
{\cal L}_\textrm{4-{\rm boson}}\uprm{\textrm{super-FTPR}}
+{\cal L}_\textrm{2-{\rm fermion}}\uprm{\textrm{super-FTPR}}
+{\cal L}_\textrm{4-{\rm fermion}}\uprm{\textrm{super-FTPR}},
\\ \label{BosFTPR}
{\cal L}_\textrm{4-{\rm boson}}\uprm{\textrm{super-FTPR}}&=
\co_{\rm FTPR}=
\frac{1}{\bar\phi}
\left[
\left(\nabla^2\right)^2-\frac{3}{2r^2}\nabla^2+\frac{4}{r^2}\partial_t^2
-\frac{9}{16r^4}
\right]
\frac{1}{\phi},
\\
{\cal L}_\textrm{2-{\rm fermion}}\uprm{\textrm{super-FTPR}} &=
-\bar\psi^\alpha
\left[\left(\nabla^2-\frac{3i}{r}\partial_t+\frac{2}{r^2}\right)\frac{1}{\bar\phi^2}\right]
\left[\left(\gamma^\mu_{\alpha\beta}\nabla_\mu + \frac{i}{r}\gamma^t_{\alpha\beta} \right)\frac{1}{\phi^2}\right]\psi^\beta,
\\
{\cal L}_\textrm{4-{\rm fermion}}\uprm{\textrm{super-FTPR}} &= 
-\frac{\bar\psi_\beta\bar\psi^\beta}{\bar\phi^3}\left(\nabla^2-\frac{1}{4r^2}\right)\frac{\psi^\alpha\psi_\alpha}{\phi^3}.
\end{align}
Here $\gg\sqd$ is the full three-dimensional Laplacian, not the Laplacian on the spatial directions only.


\heading{Sign constraint}

The four-derivative, zero-fermion term in the flat-space classical action comes entirely from the FTPR term \rr{FTPRFlatSpace}.
It has been pointed out     \cite{Adams:2006sv} that such a term can only appear with a positive sign in the effective action for a massless field.
A negative sign would give rise to superluminal signal propagation, as well as unitarity violation in moduli scattering, within the
regime of validity of the effective theory.  When we calculate the spectrum, we will see that the positivity of the coefficient 
(which we shall call $\alpha$) shall translate directly into a negative sign for the coefficient of the leading large-$J$ correction
to the dimension of the lowest unprotected scalar operator of large $R$-charge $J$.

\heading{Global symmetries}

In table \ref{SymmetryTable},
we present the action of the global symmetries on the fields of the UV description and on the $\phi, \psi$ fields
of the moduli space of the $X$ branch.  Note that the $\phi$-number and $\psi$-number symmetries are \textit{separately} conserved
as exact symmetries in the moduli space effective theory, not merely accidental symmetries.  These separate boson- and fermion-number conservation laws simplify the classification of states and operators in the large-$J$ effective theory to a considerable extent.
\begin{table}
  \begin{center}
  \begin{tabular}{c|c|c|c|c|c}
     & $U(1)_X$ & $U(1)_{YZ}$ & $U(1)_R$ & $U(1)_\phi$ & $U(1)_\psi$  \\
    \hline\hline
    $W$ &  $0$ & $0$  & $+2$ & $+2$  &  $-2$   \\ \hline
    $Q$ & $0$ & $0$ &$-1$ & $-1$ & $+1$   \\ \hline
    $\bar{Q}$ & $0$ & $0$ & $+1$ & $+1$ & $-1$   \\ \hline
    $X$ & $+1$ & $0$ & $+2/3$ &  $+4/3$ & $0$  \\ \hline
    $Y$ & $-1/2$ & $+1$ & $+2/3$ & $+1/3$  & $-1$  \\ \hline
    $Z$ & $-1/2$ & $-1$ & $+2/3$ & $+1/3$ & $-1$ \\ \hline
    $\phi$ & $+3/4$ & $0$ & $+1/2$ & $+1$ & $0$ \\ \hline
    $\psi$ & $+3/4$ & $0$ & $-1/2$ &$0$ & $+1$
  \end{tabular}
  \caption{$R$ and non-$R$ global charges.  The charge assignments in the effective theory are $J_\phi = {2\over 3} J_X + J_R$ 
  and $J_ \psi = {2\over 3} J_X - J_R $.  The $U(1)_{YZ}$ symmetry acts trivially on all light states on the $X$ branch.  The 
  fermion-number symmetry $U(1)\ll\psi$ is unbroken even when $\phi$ gets an expectation value, and organizes Feynman rules
  in large-$J$ states.} \label{SymmetryTable}
  \end{center}
\end{table}

\section{Quantization of the effective $X$ branch theory}\label{XBranchQuant}

We now derive the Feynman rules for the quantization of the effective theory of the $X$ branch.  Our approach is the standard
approach to the quantization of an effective field theory.  We have a double hierarchy $E\ll{\rm IR} \muchlessthan \L \muchlessthan
M\ll{\rm UV}$, where $\Lambda$ is the cutoff (of unspecified form) and $M\ll{\rm UV}$ is the ultraviolet scale set by the "vev" of
$|\phi|\sqd$.  We will be working in finite volume, so the "vev" is not truly a vacuum expectation value; however we shall refer to it
as a "vev" anyway informally.  Later on we will comment on the physically and mathematically precise meaning of the vev in
the sense we use it.  For now, it is sufficient to refer to it by its operational meaning: We define the path integral by dividing $\Phi
\equiv X\uu{3\over 4}$ into a "vev" $\Phi\ll 0$ and a fluctuation $f$, and path integrate over $f$ in the usual way, imposing
Feynman boundary conditions on it.

\heading{Vev and fluctuations}

\begin{sloppypar}
We begin by defining
\begin{align}
	\begin{split}
X \equiv \Phi \uu{4\over 3}\ , \llsk\llsk \Phi = X\uu{3\over 4} 
	\end{split}
\end{align}
and decomposing the bosonic component $\phi \equiv \Phi\big |\ll{\th = \thb = 0}$ into
\begin{align}
	\begin{split}
\phi \equiv \phi\ll 0 + F\ \equiv e^{it/2}(\varphi_0 +f), 
	\end{split}
\end{align}
where $F$ is presumed to satisfy the free-field equation of motion $\nabla^2 \phi = + {1\over 8} {\tt Ric}\ll 3 \phi 
 = {1\over {4r\sqd}} \cc 
\phi$, and $\varphi_0$ is constant.  We then decompose the FTPR term into vev and fluctuations, retaining terms of four or fewer fluctuations. 
Note that substituting $\phi_0$ into $\mathcal{O}_{\text{super-FTPR}}$ will only yield zero,
so that the classical correction vanishes.
\end{sloppypar}

We now list a few vertices that will be relevant later.
By explicit computation, there are no corrections quadratic in fluctuations, modulo terms proportional to the leading-order equations
of motion.  By Lorentz invariance this must be so in flat space; by Weyl rescaling it is automatically the case on $S\uu 2 \times \IR$
as well.  There are also cubic vertices with three bosonic fluctuations, as well as two fermions and one bosonic fluctuation; however these do not contribute to the observables we will calculate and we do not list them.  In general, the FTPR vertex with $n\ll{\rm B}$ bosonic fluctuations and $n\ll {\rm F}$ fermions, scales as 
$|\phi\ll 0|\uu{-(n\ll B + n\ll {\rm F} + 2)}$, so the cubic vertices scale as $|\phi\ll 0|\uu{-5} \propto J\uu{-5/2}$
and the quartic vertices scale as $|\phi\ll 0|\uu{-6} \propto J\uu{-3}$. 
Hereafter we denote the propagation of $F$ by solid lines and
that of $\psi$ and $\bar\psi$ by dotted lines.

The 4-point bosonic vertex is
\begin{align}\begin{split}
\includegraphics[valign=c]{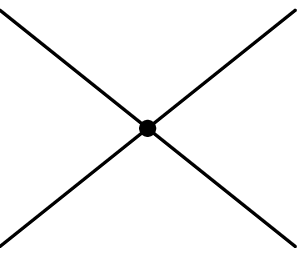}
&=
\frac{2\bar{F}^2}{{\bar\phi_0}^3}
\left[
(\nabla^2)^2-\frac{3}{2r^2}\nabla^2+\frac{4}{r^2}\partial_t^2
-\frac{9}{16r^4}
\right]
\frac{2F^2}{{\phi_0}^3}
\label{QuarticBosonVertex}
\\
&\propto J^{-3}.
\end{split}
\end{align}
The vertex with two fermions and one bosonic fluctuation is
\begin{align}\begin{split}
&
\includegraphics[valign=c]{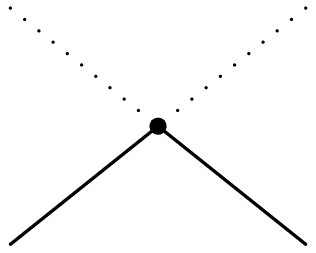}
\\
&=\displaystyle
-4\bar\psi^\alpha
\left[\left(-\partial_t^2+\nabla^2-\frac{3i}{r}\partial_t+\frac{2}{r^2}\right)\frac{\bar{F}}{\bar\phi_0^3}\right]
\left[\left(\gamma^\mu_{\alpha\beta}\nabla_\mu + \frac{i}{r}\gamma^t_{\alpha\beta} \right)\frac{F}{\phi_0^3}\right]\psi^\beta
\label{TwoFermionTwoBosonVertex}
\\&\propto J^{-3}.
\end{split}
\end{align}
and the vertex with four fermions and no bosonic fluctuations is
\begin{align}\begin{split}
\includegraphics[valign=c]{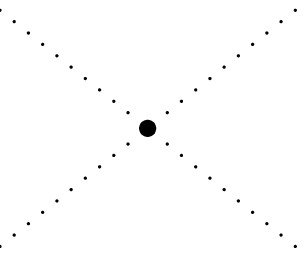}
&=\displaystyle
\frac{\bar\psi_\alpha\bar\psi^\alpha}{\bar{\phi}_0^3}
\left[
-\partial_t^2+\nabla^2-\frac{1}{4r^2}
\right]
\frac{\psi_\alpha\psi^\alpha}{\phi_0^3}
\label{FourFermionVertex}
\\&\propto J^{-3}.
\end{split}
\end{align}

As we will explain later when we give a precise definition of the "vev" $\phi\ll 0$, the $F$ field satisfies Feynman boundary conditions, 
thus has the usual Feynman propagator.  

\heading{Feynman rules and $J$-scaling of diagrams}  

Having decomposed the field into vev and fluctuations, we can most easily understand the scaling of
corrections by writing the Feynman rules for the $f$ and $\psi$ fields.
A diagram with $m$ FTPR vertices with $k\ll 1, k\ll 2, \cdots, k\ll m$ lines on each vertex, will scale as $|\phi\ll 0|\uu{- 2m - \sum k\ll i}$,
and therefore as $J\uu{- m - \hh \cc \sum k\ll i}$.

\section{Corrections to operator dimensions}

\subsection{Dynamics on $  S\uu 2\times\IR$}

Now we are going to study the theory in radial quantization, which means computing operator dimensions
as energies on  $S\uu 2$ with radius $r$, in units of ${1\over r}$.  We will focus on the lowest states
with given global charges (in particular the $R$-charge), as well as low-lying excited states above the lowest.  We will see 
that the lowest state is described by a classical solution with a particular symmetry, and that the fluctuations around
the classical solution are weakly coupled when the global charges are large.  Thus the quantum properties of the
lowest state and states of low excitation number above the lowest, are calculable in a perturbation series with
expansion parameter ${1\over J}$.

\heading{Classical solutions with lowest energy for a given global charge}

There is a particular family of classical solutions on $S\uu 2 \times \IR$ that 
saturates the lower bound on the energy for a given $R$-charge, $E \geq J\ll R / r$, where $r$ is the radius of the sphere.

This solution exists regardless of the form of the terms in the effective action for $X$.  This follows from a general fact
in classical mechanics: The lowest classical solution with a given value of a conserved charge $J$, always preserves 
a "helical" symmetry, \it i.e., \rm a combined symmetry under a time translation and action of the conserved charge $J$ by
Poisson brackets.  Furthermore, the angular frequency of the global symmetry action is given by $\o\equiv {{dE}\over{dJ}}$.
In the case where the global symmetry is an $R$-charge, the lowest classical solution
with a given $J\ll R$ is invariant under a combined time translation and $R$-symmetry rotation, and the angular frequency
of the $R$-symmetry rotation is exactly $1/r$ for any value of the amplitude.  That is, the lowest classical solution(s) carrying
a given value of the $R$-charge have a helical symmetry, with $X,Y,Z$ depending on time as $\exp\{2 i t / 3r\}$.

Due to the presence of the $XYZ$ superpotential, there is no such solution with more than one of $X,Y,Z$ turned on at a time,
so the lowest classical solutions carrying a given $R$-charge are simply $X = X\ll 0 \exp\{2it/3r\}$, and then two other
branches of solutions, with $X$ replaced by $Y$ or $Z$.  These branches of solutions
have $X,Y, $ and $Z$ charge equal to ${3\over 2}$ times their $R$-charge, respectively.  Due to the $S\ll 3$ symmetric group
permuting the three branches, we can ignore the $Y$ and $Z$ branches, and focus exclusively on the properties of
the $X$ branch.

Unlike the values of of $E$ and $\o$ for a given $J\ll R$, which are universal, the amplitude $|X\ll 0|$ of
the helical solution as a function of $J\ll R$ depends on the unknown form of the Wilsonian action.  However we can
estimate the value of $|X\ll 0|$ on $J\ll R$ using dimensional analysis.  All scale invariant bosonic terms take the form of polynomials in derivatives of $X$ and $\bar{X}$ in the numerator, dressed with the appropriate power of $|X|$ in the denominator to render the term scale invariant.  So each additional derivative (or curvature) in the numerator costs an additional power of $|X|\sqd$ (or two) in the denominator.
Thus the derivative (and curvature) expansion of the Lagrangian is also an expansion in inverse power of $|X|$, because of the
underlying conformal invariance of the theory, which we use as an input in constraining the action.  It follows that the leading term
in the effective action for $X$, which is simply the free kinetic term with conformal coupling to the Ricci scalar, controls the
leading large-$J$ asymptotics of the magnitude of $|X\ll 0|$ in the helical solution.  Thus we conclude that 
$|X\ll 0|$ is proportional to $({J\over r})\uu{3\over 4}$ at leading order, with a coefficient depending 
on the normalization of the kinetic term and corrections that are subleading at large $J$.

\subsection{Meaning of the "vev"}\label{VevMeaning}

\heading{Free-field matrix elements with a "vev" are coherent state matrix elements}

In finite volume, there is of course no such thing as spontaneous breaking of global symmetries.  This can be seen easily from
the fact that the expectation value of a charged operator in a state of definite charge, is always zero.  This statement holds
only if the state is an \rwa{exact} charge eigenstate.  However states with exactly Gaussian correlation function for charged
free fields can be constructed as coherent states.  If $a^{\dagger}\equiv(a\dag)\uu\phi$ is a creation operator for an excitation of the $\phi$
field in the $s$-wave, then the coherent state
\begin{align}
	\begin{split}
\dket{[v]} \equiv \exp\{ v\cdot a\dag\} \cc \kket{0}
	\end{split}
	\label{CoherentStateDef}
\end{align}
has the property that 
\begin{align}
	\begin{split}
a\dket{[v]} = v \cc \dket{[v]}\ ,
	\end{split}
\end{align}
and correlation functions of the oscillators in the state $\dket{[v]}$ are exactly Gaussian.  Therefore free fields $\phi, \phb$ built
from the oscillators have the property that $f\equiv \phi - \left<\phi\right>$ and $\bar{f}\equiv \phb - \left<\phb\right>$ have the same correlation functions
as the vacuum correlation functions of $\phi$ and $\phb$:
\begin{align}
	\begin{split}
\left<{[v]}\left| {\cal O}\{\phi, \phb\} \right| {[v]} \right>
= \bbra 0 {\cal O}\{ f, \bar{f} \} \kket 0\ ,
	\end{split}
\end{align}
where we have defined
\begin{align}
	\begin{split}
\kket{[v]} \equiv {\cal N}\ll v\uu{-\hh} \cc \dket{[v]}\ , \qquad
 {\cal N}\ll v\equiv \nbdbra{[v]} \dket{[v]} = \exp\{+ |v|\sqd\}\ .
	\end{split}
\end{align}

It remains to compare the expectation values in a large-$J$ eigenstate with expectation values in a coherent state\footnote{This calculation was developed by one of the authors (SH) with Ian Swanson, and used to estimate corrections to the energies
of rotating relativistic strings \cite{Hellerman:2013kba}.  
}
and to show that the latter approximates the former in the large-$J$ limit,
with calculable corrections.

Using $\kket{\cc J} = {1\over{\sqrt{J!}}} (a\dag)\uu J \cc\kket {0}$, the definition \rr{CoherentStateDef} can be written
\begin{align}
	\begin{split}
\dket{[v]} = \sum\ll J {{v\uu J}\over{\sqrt{J!}}} \cc \kket {\cc J} 
	\end{split}
\end{align}
and inverted to give
\begin{align}
	\begin{split}
\kket {\cc J}  = (2\pi i)\uu{-1} \cc \oint \cc {{dv}\over{v\uu {J+1}}} \cc \dket{[v]}\ .
	\end{split}
\end{align}
(Here, $J$ denotes $J\ll \phi$, the charge of the $\phi$-oscillator). 

This state has exactly Gaussian correlators, with a connected two-point function identical to that of the (uncharged) vacuum.  It follows that
the relation between the vacuum and coherent-state two-point function is simply a shift of the one-point functions, by a free classical solution.
We can therefore use Feynman diagrams in a "background" given by the classical solution represented by the coherent state expectation value,
to calculate arbitrary free-field correlation functions in the coherent state.  So the usual Feynman diagrammatic perturbation theory with a "vev" given by
a nontrivial free classical solution for
the scalar field, is simply a way of doing time-ordered perturbation theory in the coherent state in finite volume.
This is relevant to the large-$J$ expansion for definite-$J$ matrix elements, because as we will now see, large-$J$ matrix elements
in charged Fock states are approximated at leading order by matrix elements in the corresponding coherent state.

\heading{Relationship between Fock states and coherent states}

A consequence of this representation is that expectation values for Fock states are approximated at leading order in $J$ by expectation
values in coherent states, up to (calculable) subleading large-$J$ corrections.  To see this concretely, we need the following facts:

\bi
\item{The definite-$J$ Fock matrix element is given by a double contour integral of coherent-state matrix elements;}
\item{The double contour integral for a neutral operator can be evaluated by saddle point;}
\item{Fluctuation corrections to the saddle-point approximation are suppressed by powers of $J$; and}
\item{The leading saddle-point approximation is simply given by the coherent-state matrix element in the coherent state where the expectation value
of the charge, is $J$.}
\ei

First we write the expectation value $A\upp{\cal O}[J] \equiv \bbra {J} {\cal O} \cc \kket{J}$ in the state $J$ as
a double contour integral,
\begin{align}
	\begin{split}
A\upp{\cal O}[J] = (2\pi)\uu{-2} \cc \oint \cc \oint\cc {{dw}\over{w\uu{J+1}}} \cc {{ dv}\over{v\uu {J+1}}}
\cc\dbra{[w]} {\cal O} \cc \dket{[v]} 
\label{Higher big-J corrections}
	\end{split}
\end{align}

One combination of the two integrals simply projects
onto operators ${\cal O}$ that commute with $\hat{J}$.
Assume WLOG that ${\cal O}$ carries a definite charge,
\it i.e., \rm $[\hat{J}, {\cal O}] = J\ll {\cal O} \cc {\cal O}$.
If $J\ll{\cal O} \neq 0$, then clearly its expectation value
in Fock states must vanish.  The first of the two contour
integrals simply
implements the projection that causes the Fock state
expectation value to vanish.

If ${\cal O}$ is uncharged, the remaining contour integral
is nonzero, and can be evaluated by saddle point
when $J$ is large, with fluctuation corrections
that can be calculated as a series in ${1\over J}$.  Define $F\upp{{\cal O}}[J_{\rm cl}]$ as the
expectation value of an uncharged operator ${\cal O}$ in
a coherent state of classical charge equal to 
$J$:
\begin{align}
	\begin{split}
F\upp{{\cal O}}[J] \equiv \bbra{[w]} {\cal O} \cc \kket{[v]} 
\biggl |\ll{J \equiv w\cdot v}\ ,
	\end{split}
\end{align}
Then the Fock expectation value $A\upp{\cal O}[J]$ is given by
\begin{align}
	\begin{split}
A\upp{\cal O}[J] = 
\sum\ll{m,n\geq 0} \cc \hh \cc {\cal R}\ll{mn} J\uu m \cc \bigg ( {d\over{dJ}}
   \bigg ) \uu n \cc F\upp{{\cal O}}[J] \ ,
	\end{split}
\end{align}
where the leading coefficient ${\cal R}_{00}$ is $1$, and all the 
other coefficients are given by the
generating function
\begin{align}
	\begin{split}
\sum\ll{m,n\geq 0} {\cal R}\ll{mn}x\uu m y\uu n = e^{-xy}\cc (1 + y)\uu x\ .
	\end{split}
\end{align}
Note that ${\cal R}\ll{mn} = 0$ unless $m \leq n/2$, so there
are only a finite number of nonzero terms at a given
order in $J$.  Concretely, if we expand
\begin{align}
	\begin{split}
A\upp{\cal O}[J] = \sum\ll{k\geq 0} A\upp{\cal O}\ll k[J]
	\end{split}
\end{align}
where $A\upp{\cal O}\ll k[J]$ is the relative-order $J\uu{-k}$ contribution to the Fock-state expectation value, then
\begin{align}
	\begin{split}
A\upp{\cal O}\ll k[J] = \sum\ll{n - m = k} {\cal R}\ll{mn}J\uu m 
 \cc \bigg ( {d\over{dJ}}
   \bigg ) \uu n \cc F\upp{{\cal O}}[J]\ ,
	\end{split}
\end{align}
and the first few contributions are
\begin{align}
\begin{split}
A\ll 0[J] &= F[J]\ ,\\ 
A\ll 1[J] &= - \hh J \cc F\prpr[J]\ ,  \\ 
A\ll 2[J] &= + {1\over 8} \cc J\sqd F\uu{\prime\prime\prime\prime}[J] + {1\over 3} \cc J F\prprpr[J]\ .
\end{split}
\end{align}
As expected, the leading approximation $A\ll 0[J]$ is simply equal to the coherent-state expectation value.

\heading{Conical deficit and $\phi$-charge quantization}

The change of variables $\phi \equiv X\uu{3\over 4}$ is
well-behaved at large values of $X$ (compared to the infrared
scale) but singular at the origin.  The classical helical solution
never comes near the origin of field space, nor do fixed-energy
perturbations of the helical solution in the limit of large $J$.  So one
would expect the singularity of the change of variables to be
irrelevant in large-$J$ perturbation theory.

On the one hand, this expectation is entirely accurate,
in the sense that the details of the "resolution" of
the singularity are indeed irrelevant to all orders in the $1/J$
expansion.  Any two physically well-defined resolutions of the singularity, must necessarily correspond to different Hamiltonians
$H\ll{1,2}$ that modify the moduli space effective action
in a neighborhood of $\phi$-field space of size $M\uu{+\hh}$
(equivalently, a neighborhood of $X$-field space of size $M\uu{+{3\over 4}}$), where $M$ is some ultraviolet scale.  If the correction terms scale like $ M\uu{k\over 2} / |\phi|\uu k$ at long distances
in field space, then the corresponding large-$J$ corrections
go as $\left ( {M / J} \right )\uu{k\over 2}$.  If the
two resolutions of the geometry are both
\it exactly \rm conical outside of a
a region of field space $|\phi| < M\uu\hh$, then the corrections
to observables from the modified geometry vanish to
all orders in $J$.  This is the precise sense in which the singularity
at the origin is "irrelevant" for large-$J$ physics: At large
$J$, the field doesn't live at the origin or anywhere near it.

However the conical deficit is a property of the geometry that
is visible asymptotically, and the effective theory should
know about all properties of the moduli space 
geometry where the vev is large compared to the infrared scale.
The quantization rule for $\phi$-charge is precisely the property
of the quantum effective theory that encodes the conical deficit
at large vev.  For purposes of deriving a large-$J$ asymptotic expansion 
in the effective field theory, one may simply take $J\ll X$ to be a multiple of $3$ , in which case the number of $\phi$-excitations
is an integer, and in particular a multiple of 4.  

To verify that the only effect of the conical deficit is to alter the quantization rule, one can simply repeat any calculation
in the $\phi$ effective theory, in terms of a logarithmic superfield defined as $L \equiv {\tt ln}(\Phi)$.  In terms of $L$,
the only effect of the conical deficit is to alter the periodicity of the imaginary part of $L$; otherwise the Lagrangian
is completely unaffected by the deficit.  We conclude that the conical deficit has no effect on the energy spectrum
to any order in perturbation theory, so long as the classical solution uniformly satisfies $|\phi|\sqd \muchgreaterthan E\ll{\rm IR}$.

\subsection{BPS property and vanishing of the vacuum correction}

\heading{The classical energy of the large-$J$ ground state}

\begin{sloppypar}
First, as a consistency check, we shall examine the energies of the BPS states, at the classical and one-loop level.  By general multiplet-shortening
arguments \cite{Dobrev:1985qv,Dobrev:1985vh,Minwalla:1997ka,Kinney:2005ej}, these energies must remain uncorrected, and equal to the $R$-charge of the state.  However even
at the classical level, it is not immediately apparent that the 
super-FTPR term leaves the energies of the BPS states unaffected at the classical level.
The term is a sum of many contributions with particular coefficients, none of which individually vanishes for the helical ground state classical solution.
Nonetheless the sum of the terms in the FTPR expression 
\rr{BosFTPR} does indeed combine to give zero when evaluated on the helical solution:
\begin{align}
\begin{split}
\co_{\rm FTPR}\uprm{bosonic} =
\frac{1}{\bar\phi}
\left[
\left(\nabla^2\right)^2-\frac{3}{2r^2}\nabla^2+\frac{4}{r^2}\partial_t^2
-\frac{9}{16r^4}
\right]
\frac{1}{\phi} = 0 \ ,
\end{split}
\end{align}
for any spherically homogeneous helical solution with frequency ${1\over{2r}}$, $ \phi = \exp\{i t / (2r)\} \cc \phi\ll 0$.
This gives us some confidence in the applicability of the moduli space effective action to compute energies consistently.
\end{sloppypar}

Next, we shall compute the 
one-loop energies of the ground states, as well as semiclassical and one-loop energies of first-excited states as a consistency check, to build further confidence in our methods.

\heading{One-loop energy of the large-$J$ ground state}

We now check the one-loop energy of the large-$J$ ground state, by expanding the action around the helical solution to quadratic order in fluctuations, and
summing $\pm \o$ over bosonic and fermionic fluctuations with frequency $\o$, with the sign appropriate to the statistics.  At the free level, the
 bosonic and fermionic fluctuations are paired at each frequency, and thus their contributions to the vacuum energy cancel mode by mode.  The
super-FTPR term could in principle have corrected the frequencies at order $J\uu{-2}$ but it does not: As noted in 
section \ref{XBranchQuant}, the super-FTPR term, when expanded around the helical solution, contains no pieces quadratic in fermions or in bosonic fluctuations,
and thus the energy is automatically uncorrected at absolute order $J\uu{-2}$ (which is relative order $J\uu{-3}$) even without any further nontrivial Bose-Fermi cancellation.
 So we see that the energy
of the BPS ground state is therefore uncorrected up to and including order $J\uu {-2}$, as it must be to all orders in $J$.  A nontrivial test of the large-$J$ expansion
would be to verify the cancellation of the correction to all orders in the loop expansion.  It may be that some type of superfield formalism adapted to quantization about the helical
ground state would make such cancellations more transparent.

\subsection{Semi-short property of the $s$-wave one-particle state}

Next we turn to the computation of first-excited energies at large $J$.  The lowest state above the large-$J$ ground state with
the same $U(1)$ quantum numbers, is the state with
an additional $\phi$ excitation and $\phb$ excitation, both in the $\ell = 0$ mode, \it i.e., \rm the $s$-wave.  At the free level, each has frequency $\o = {1\over{2r}}$, and
we have seen that the frequency is unaffected by the interaction term up to and including order $J\uu{-2}$.  So up to and including order $J\uu{-2}$, the
energy of the first excited state is simply $J+1$.  

Since this state is not a BPS chiral primary, one might be interested to calculate corrections to its energy in the large-$J$ expansion.  However the
one-loop correction actually vanishes.

Heuristically, the first-excited state can be thought of as obtained by shifting the $\phi$ charge
of the vacuum from $J\to J+1$, and then cancelling it
by adding a single quantum of $\phb$ in the $s$-wave:
\bbb
\phi = g \cc \exp\left(-\frac{ i t}{2r}\right) \ , \llsk\llsk \phb = g\st \cc \exp\left(+ \frac{i t}{2r}\right)\ .
\eee
Note that this linearized solution does not preserve the helical symmetry
of the ground state; it is $\phb$ rather than $\phi$ that has a positive-energy
mode excited.  The one-loop correction to the energy of the state
coming from a quartic vertex, can be seen to cancel
explicitly.  The one nontrivial aspect of this
cancellation is the operator ordering of the quartic
term: Since the vev corresponding to the
coherent BPS state breaks time-reversal invariance
spontaneously, the ordering of operators is not
simply described in terms of time-ordered terms.  The
more convenient description is in terms of normal-ordered
operators in the Hilbert space on the sphere: All operators
appearing have at least two $\phb$-multiplet annihilation oscillators ordered to the right, and thus
the perturbing Hamiltonian does not affect the energy
of the semishort state, which has only a single $\phb$ excitation.

The existence of semishort states with the appropriate charges is visible at the level of the superconformal
index; we have included an expression for the index in the Appendix, as well as its expansion to several orders,
so that the reader may see the agreement for herself.
It is interesting to note that the semi-short states of the $X$-branch persist down to $J=0$: The "moment map" operator is semi-short on
general grounds, because it the superconformal primary whose descendant is
the $U(1)\ll X$ current \cite{Green:2010da,Cordova:2016xhm}. This operator
can be thought of as the conformal K\"ahler potential itself for the effective theory of the $X$ branch,
namely $K \propto (X\bar{X})^{{3\over 4}}$.  This representation in terms of the $X$ field is not a well-defined, controlled operator
for general purposes, but this expression is well-defined and precise in matrix elements between 
large-$J$ states.

The one-particle states with nonzero spin are also in semi-short representations at the free-field level.  At the interacting level,
 it is easy to prove in many cases that the semi-short property persists, because there are no states with the
 appropriate angular momentum, $U(1)\ll R$, and $U(1)\ll X$ quantum numbers to fill out a full long representation at weak
 but finite coupling.  For instance, the vector states obtained by acting on the BPS vacuum $\phi\uu{J+1}$ with the $\ell = 1$ modes
 of the $\phb$ field, can be shown to be protected by such an argument.  This prediction is also verified by the superconformal index.

\subsection{Correction to the two-particle energies}\label{quant}
\heading{Semi-short property means no disconnected diagrams}

Say we calculate the energy correction to
the state with two $\bar\phi$s on top of $\phi^J$.
By expanding in VEV and fluctuations, two diagrams to consider at order $J^{-3}$ are as follows:
\begin{eqnarray}\label{4p}
\includegraphics[valign=c]{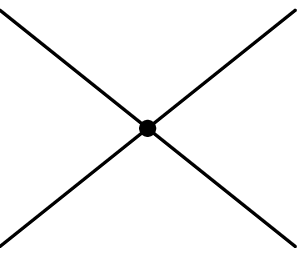}
\end{eqnarray}
and
\begin{eqnarray}
\includegraphics[valign=c]{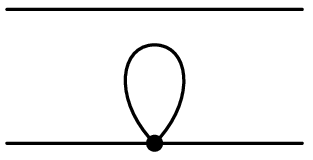}
\end{eqnarray}
Incidentally, we know from the argument in the last subsection
that this diagram should vanish:
\begin{eqnarray}
\includegraphics[valign=c]{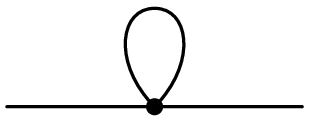}
\label{twop}
\end{eqnarray}
Note that these Feynman diagrams with loops in them have
scheme dependence, {\it i.e.},
how you regularize and renormalize loop integrals --
once we choose one scheme that is compatible with
supersymmetry on $S^2\times \mathbb{R}$,
the expression is meaningful, and the diagram
(\ref{twop}) gives exactly zero. Hence the only contribution to the energy correction
at order $J^{-3}$ is the diagram (\ref{4p}).

The above argument holds even when
some of the $\bar\phi$s are changed into $\bar\psi$, the $\bar{Q}$-descendant of $\phb$. The only diagram
that contributes to the energy correction to the state
$\bar\psi\bar\phi\Ket{\phi^J}$
is
\begin{eqnarray}
\includegraphics[valign=c]{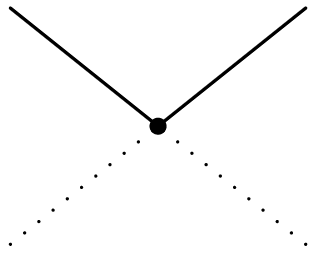}
\end{eqnarray}
while
for $\bar\psi\bar\psi\Ket{\phi^J}$
this is 
\begin{eqnarray}
\includegraphics[valign=c]{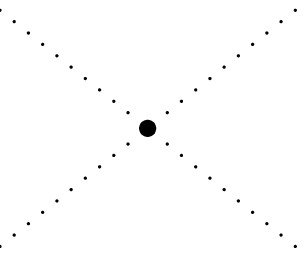}
\end{eqnarray}

\heading{Two-particle states energy correction}

Now let us calculate the energy correction to the state
$\bar\phi\bar\phi\Ket{\phi^J}$ by expanding in VEV
and fluctuations. 
Here note that we only have to care about spatially uniform
field because of the argument in the last subsection.
Truncating $f$ to $s$-waves accordingly,
we get the Lagrangian density for the fluctuation $f(t)$,
\begin{align}\begin{split}
\mathcal{L}&=\mathcal{L}_0+\alpha\mathcal{L}_{\rm int} \label{bosonicint} \\
\mathcal{L}_0&=\dot{\bar{F}}\dot{F}-\frac{1}{4}F^2
=\dot{\bar{f}}\dot{f}+\frac{i}{2}(\dot{\bar{f}}f-\bar f\dot f)\\
\mathcal{L}_{\rm int} &= -\frac{24}{|\varphi_0 |^6}\bar f^2\dot f^2,
\end{split}
\end{align}
Above we have for simplicity set the radius $r$ of the sphere to $1$, as we shall continue to do
in the rest of this section.
Dots represent derivative with respect to $t$
and the radius of $S^2$ is set equal to unity.
We derive the Hamiltonian ($4\pi $ times the Hamiltonian density)
of the system from this Lagrangian.
The conjugate momentum in terms of $f$ and $\bar f$ is
$\Pi:=\dot{\bar{f}}-\frac{i}{2}\bar f$ and 
$\bar\Pi:=\dot f+\frac{i}{2} f$, respectively.
\begin{align}
\begin{split}
{H}&={H}_0+\alpha{H}_{\rm int}\\
{H}_0&=4\pi
\left(\Pi+\frac{i}{2}\bar f\right)\left(\bar\Pi+\frac{i}{2}f\right)
\\
{H}_{\rm int} &= 4\pi\times \frac{24}{|\varphi_0 |^6}\bar f^2\dot f^2.
\end{split}
\end{align}
As always we define creation and annihilation operators as
\begin{align}\begin{split}\label{Creation}
a^\dagger&=\sqrt{4\pi}\left(\Pi+\frac{i}{2}\bar{f}\right), \qquad 
a=\sqrt{4\pi}\left(\bar{\Pi}-\frac{i}{2}f\right),\\
b^\dagger&=\sqrt{4\pi}\left(\bar\Pi+\frac{i}{2}f\right), \qquad 
b=\sqrt{4\pi}\left(\Pi-\frac{i}{2}\bar{f}\right),
\end{split}\end{align}
and ${H}_{\rm int}$ becomes
\begin{equation}
-\frac{1}{|\varphi_0|^6}\times\frac{6}{\pi}\times
(a^\dagger-b)(a^\dagger-b)aa.
\label{OperatorPerturbation}
\end{equation}
We then evaluate the energy correction to the state
$a^\dagger a^\dagger\Ket{0}$, which corresponds to $\bar\phi\bar\phi
\Ket{\phi^J}$:
\begin{equation}
\Delta E=-\frac{6\alpha}{|\varphi_0|^6\pi}\frac{\Braket{0|aa\left[
(a^\dagger-b)(a^\dagger-b)aa\right]a^\dagger a^\dagger
|0}}{\Braket{0|aaa^\dagger a^\dagger|0}}
=-\frac{12\alpha}{|\varphi_0|^6\pi}.
\label{OperatorPerturbationMatrixElement}
\end{equation}
Likewise, the energy correction to $\bar\psi\bar\phi
\Ket{\phi^J}$ and $\bar\psi\bar\psi
\Ket{\phi^J}$ can be evaluated and gives the same number
as that of $\bar\phi\bar\phi
\Ket{\phi^J}$, as it should be due to supersymmetry.
Detailed calculations are given in the Appendix \ref{2b2f4f}.

Note that the form of the operator perturbation is normal-ordered rather than time-ordered.  This is
a consistency condition for supersymmetry, and it can be derived directly from the necessity of the existence of
a set of operators implementing the ${\cal N} = 2$ superconformal algebtra; in fact, it can be 
seen just from the consistency of a smaller algebra generated by half the generators, namely
those preserving the BPS states $X\uu J$.  In appendix \ref{SemishortSuperalgebra}, we demonstrate how the closure of the operator algebra in the interacting theory directly dictates the form of the operator perturbation of the Hamiltonian in a toy model,
obtained by truncating the $\phb$ multiplet down to its zero mode on the sphere.

In the matrix element \rr{OperatorPerturbationMatrixElement}, note that disconnected contributions coming from
vacuum bubbles and propagator corrections
are absent; the vanishing of these contributions follow from the nonrenormalization of the vacuum energy and semishort
one-particle energy, respectively.

We also draw the attention to the relative negative sign between the parameter $\a$ in the Lagrangian, and the
first correction to the non-BPS operator dimension.  A positive value for $\a$ implies a negative anomalous dimension,
and vice versa.  As noted in section \ref{XBranchEFT}, the sign of $\a$ must be positive due to a superluminality
constraint \cite{Adams:2006sv} and thus the order $J\uu{-3}$ contribution to the anomalous dimension is negative.

We have also calculated, in appendix \ref{2b2f4f}, the energy shifts of the one-boson, one-fermion state with the fermion in the
$\ell = \hh$ sector, and also the two-fermion state with both fermions in the $\ell = \hh$ mode.  These are
the $\bar{Q}$ and $\bar{Q}\sqd$-descendants of the primary state, respectively.  
The first-order interaction
contributions to the anomalous dimensions are the same as for the two-boson state, which nontrivially checks that our
formalism implements superconformal symmetry consistently.

\section{Operator algebras and the semishort spectrum}

In this section we return to the question of the semishort energy (non)correction.  We have seen explicitly that the one loop correction
to the scalar semishort energy cancels, and there is an algebraic argument that the scalar semishort energy is
unrenormalized to all orders in large-$J$
perturbation theory: The energy can only change if the semishort joins with other states to form a long multiplet, and there
are no states with the correct quantum numbers, that are near energy $\D = J\ll{\rm R} + 1$ at large $J$, to fill out a long
multiplet.  This establishes that there is a scalar semishort at sufficiently large $J$.  In this section we will see that this fact
has consequences at \rwa{low} values of $J$ due to the structure of the operator algebra: Scalar semishorts 
form a module over the chiral ring, and as a result associativity will relate semishorts at high $J$ to those at low $J$.

\heading{Nonsingularity of certain OPE structure functions}

In any unitary ${\cal N} = 2$ SCFT in three dimensions, all superconformal primary operators must satisfy \cite{Dobrev:1985qv,Dobrev:1985vh,Minwalla:1997ka,Kinney:2005ej}, 
\bbb
\Delta \geq R + s\uu 3.
\een{BPSBound}
with the inequality saturated if and only if the state is annihilated by $\Qbar\ll\downarrow$:
\bbb
\Delta\ll\co = R\ll\co + s\uu 3\ll\co
\quad \Leftrightarrow\quad
  \Qbar\ll\downarrow \co = 0\ .
\een{BPSOp}
(In fact, we need not even assume $\co$ be fully superconformal
primary; we need only assume it is annihilated by the
energy- and R-charge-lowering supercharge $S\ll\uparrow \equiv (\bar{Q}\ll\downarrow)\dag$ conjugate to $\bar{Q}\ll\downarrow$.)

By acting with an $SO(3)$ rotation on the supercharge we can change the axis of
the spin, and by exchanging $Q$ with $\Qbar$ we can send $R\to - R$, so in general we have the BPS bound
\bbb
\Delta \geq |R| + s\ ,
\een{GeneralBPSBound}
where $s$ is the total spin.  This bound is saturated if and only if the operator is annihilated by one of the four energy
raising supercharges $Q\ll\a, \Qbar\ll\a$.

For instance, an element of the chiral ring, \it i.e., \rm a scalar superconformal primary operator $\co$ satisfying 
\begin{align}
\begin{split}
\Qbar_\alpha \co= 0\ ,
\end{split}
\end{align}
has dimension equal to its $R$-charge
\begin{align}
\begin{split}
R= \D\ .
\end{split}
\end{align}
A scalar semi-short operator $\co\ll{\rm SSS}$, \it i.e. \rm a scalar superconformal primary satisfying $\Qbar\sqd \co\ll{\rm SSS} = 0$ but
not $\Qbar\co\ll{\rm SSS} =0$, does not saturate the bound
\rr{BPSOp}, but its $\Qbar\ll{\uparrow}$ superpartner $\co\ll{\rm SSSSP}\uprm{\uparrow}$
is annihilated by $\Qbar\ll\downarrow$ and also $\bar{S}\ll\uparrow$, and therefore
does saturate it:
\bbb
\D\ll{\rm SSSSP} = R\ll{\rm SSS} + {3\over 2}\ , \llsk\llsk \D\ll{\rm SSS} = R\ll{\rm SSS} + 1\ .
\een{SSSDimension}

The scalar chiral primaries have a ring structure because their OPE is automatically nonsingular \cite{Lerche:1989uy,Cachazo:2002ry}.
The argument follows immediately from rotational invariance, scale invariance,
and the BPS formula \rr{BPSOp}.

If $\co\ll {1,2}$ are two chiral ring elements, then their dimensions and R-charges satisfy $\Delta\ll{1,2} = R\ll{1,2}$.  Their
OPE is of the form
\begin{align}
\begin{split}
\co\ll 1(\s) \co\ll 2(0) = \sum\ll i f\ll i(\s) \cc \co\ll i(0)\ ,
\end{split}
\end{align}
where $\co\ll i$ are operators of R-charge $R\ll i = R\ll 1 + R\ll 2$, dimension $\Delta\ll i$, spin $s\ll i$, 
and in particular third component
of spin equal to $s\ll i\uu 3$.  
The function $f\ll i$ has $\s$-scaling $\g\ll i \equiv \D\ll i - \D\ll 1 - \D\ll 2 = \D\ll i - R\ll 1 - R\ll 2$.  So by \rr{GeneralBPSBound}
we have $\g\ll i \geq s\ll i$, and so all the structure functions $f\ll i$ vanish at $\s =  0$ unless $\co\ll i$ is a scalar.  In the
latter case the structure function has a finite limit, the OPE is nonsingular, and the product of two BPS scalar primaries at coincident points
defines an associative multiplication which is the multiplicative structure of the chiral ring.

Now we consider the operator product of a BPS scalar $\co\ll{{\rm BPS}}$ of R-charge $R\ll{{\rm BPS}}$ with a scalar semi-short
${\cal O}\ll{{\rm SSS}}$ with $R$-charge $R\ll{{\rm SSS}}$.  The dimensions of the operators are
$\D\ll{{\rm BPS}} = R\ll{{\rm BPS}}$ and $\D\ll{{\rm SSS}} = R\ll{{\rm SSS}} + 1$.  In this case, there can be singular
terms in the OPE, such as
\begin{align}
\begin{split}
\co\ll{{\rm BPS}}(\s)  {\cal O}\ll{{\rm SSS}}(0) \sim |\s|\uu{-1} \cc \co\ll{{\rm BPS}}\pr(0) + ({\rm less~singular})
\end{split}
\end{align}
where $ \co\ll{{\rm BPS}}\pr$ is a BPS scalar primary of $\D\pr\ll{{\rm BPS}} = R\pr\ll{{\rm BPS}} = 
R\ll{{\rm BPS}} + R\ll{{\rm SSS}}$.  There is also the nonsingular term
\bbb
\co\ll{{\rm BPS}}(\s)  {\cal O}\ll{{\rm SSS}}(0) \ni \cdots + {\cal O}\ll{{\rm SSS}}\pr(0)+ \cdots
\een{ModuleRepTerm}
which will be of principal interest to us in this section.

 We would like to establish that this smooth and nonvanishing term \rr{ModuleRepTerm} in the OPE defines an associative multiplication of
 the chiral ring on the scalar semi-shorts.  We cannot draw this conclusion directly from the OPE above, because the nonsingular terms in a generic OPE
 are not in general associative; only the sum of all terms, singular and not, generally satisfy associativity when taken together.  However by taking
 a $\Qbar$-descendant, we can establish associativity of \rr{ModuleRepTerm} indirectly: By taking the $\Qbar\ll\uparrow$
 descendant, we can define an associative action of the chiral ring on the superpartners $\co\ll{{\rm SSSSP}}\uprm{\uparrow} \equiv \Qbar\ll\uparrow \cdot \co\ll{{\rm SSS}}$ of
scalar semi-shorts.

Both $\co\ll{\rm BPS}$ and $\co\uu\a\ll{\rm SSSSP} \equiv \Qbar\uu\a\cdot \co\ll{\rm SSS}$ satisfy the BPS
formula \rr{BPSOp},
 so by the same arguments as above, any structure functions $f\ll i$ appearing in their OPE
must scale as $|\s|\uu\g$, with the exponent $\g$ defined as
\begin{align}
\begin{split}
\g\equiv \Delta\ll{\rm RHS} -  \Delta\ll{\rm BPS} -  \Delta\ll{\rm SSSSP} =  \Delta\ll{\rm RHS} - R\ll{\rm BPS} - R\ll{\rm SSS} - {3\over 2}
\end{split}
\end{align}
where $\D\ll{\rm RHS}$ and $R\ll{\rm RHS}= R\ll{\rm BPS} + R\ll{\rm SSS} + 1$ are the dimension and $R$-charge of the operator on the right hand
side of the OPE.  If the spin of the operator on the RHS is $s\ll{\rm RHS}$, then the dimension satisfies the inequality \rr{GeneralBPSBound},
\begin{align}
\begin{split}
\D\ll{\rm RHS} = R\ll{\rm BPS} + R\ll{\rm SSS} + 1 + s\ll{\rm RHS}\ .
\end{split}
\end{align}
In the case where the RHS is a scalar semishort superpartner,
so $s\ll{\rm RHS} = \hh$, so
\begin{align}
\begin{split}
\Delta\ll{\rm RHS} \geq R\ll{\rm BPS} + R\ll{\rm SSS} + {3\over 2}\ , \llsk\llsk \g \geq 0\ .
\end{split}
\end{align}
This allows for two possible Lorentz-invariant tensor structures,
\begin{align}
\begin{split}\label{TwoOPETerms}
\co\ll{{\rm BPS}}(\s)  {\cal O}\uu\a\ll{{\rm SSSSP}}(0) \ni ({\rm const}\ll {\rm A}) \cc {\cal O}\uu\a\ll{{\rm A}}(0) +
({\rm const}\ll {\rm B})\cc
 \gamma\ll\m\uu{\a\b}\cc {{\s\uu\m}\over{|\s|}}{\cal O}\uu\b\ll{{\rm B}}(0)\ .
\end{split}
\end{align}
It can be shown that the second of the two tensor structures can never appear in a parity-symmetric theory such as the $XYZ$ model.  The
result is that the OPE of a chiral ring element with the superpartner of a scalar semishort operator is of the form

\bbb
\co\ll{{\rm BPS}}(\s)  {\cal O}\uu\a\ll{{\rm SSSSP}}(0) \ni ({\rm const}\ll {\rm A}) \cc {\cal O}\uu\a\ll{{\rm A}}(0) \,
\een{ReallyOnlyOneOPETerm}
where ${\cal O}\uu\a\ll{{\rm SSSSP}}$ is the scalar sem-short superpartner and the operator ${\cal O}\uu\a\ll{{\rm A}}$ is a spin-$\hh$ operator saturating the BPS bound.  Any such operator is again
necessarily the $\bar{Q}\uu\a$ descendant of a scalar semi-short, as shown by the following argument.

By virtue of the BPS bound, all other operators on the right-hand side of \rr{ReallyOnlyOneOPETerm} vanish in the limit $\s\to 0$.  Therefore
the $\bar{Q}$-descendants of scalar semishorts form a module over the commutative ring of the chiral primaries:\footnote{In $D=4$ this
module structure was argued for in \cite{Berkooz:1995cb}.}
\bbb
\co\ll{{\rm BPS}}(\s)  {\cal O}\uu\a\ll{{\rm SSSSP}}(0) = ({\rm const}) \cc {\cal O}\uu{\a\cc\pr}\ll{{\rm SSSSP}}(0) + ({\rm vanishing~at~}\s = 0)\ .
\een{BPSWithSPSSSModuleStructure}

From this, it follows that the scalar semishorts themselves form a module over the chiral ring.  Naively this would appear to follow without any
further justification, as one expects that the OPEs of descendants are completely determined by the OPEs of primary operators.  For conformal
invariance this is indeed the case, a fact synonymous with the existence and uniqueness of the conformal blocks.  For superconformally covariant OPEs, this argument does \it not \rm
generalize; there exist multiparameter families of nontrivial,
fully superconformally invariant functions of three points in superspace \cite{Park:1999cw}.  Multiplying a supercovariant
three-point function by such a function of three copies of
superspace, yields another supercovariant three-point function.
However, in the case where one of the three operators is a 
BPS scalar primary, this issue does not arise: The identity
is the only function on three copies of superspace that is 
annihilated by $\bar{D}\ll\a$ acting on any one of the three
points.

Heuristically, we can talk directly about the action of the chiral ring on scalar semi-shorts through the combined operation
\bbb
{\cal O}\ll{\rm BPS} \times {\cal O}\ll{\rm SSS}
= {\cal O}\pr\ll{\rm SSS}  \equiv \big ( \bar{Q}\ll{\uparrow} \big )\uu{-1}
\cdot {\cal O}\ll{\rm BPS} \cdot \bar{Q}\ll{\uparrow} \cdot
{\cal O}\ll{\rm SSS}
\een{ProjectorEq}
Since the chiral ring is annihilated by $\bar{Q}\ll{\uparrow}$,
the operator ${\cal O}\ll{\rm BPS}$ does indeed formally commute through $\bar{Q}\ll{\uparrow}$, justifying the above definition.  
However, the uniqueness of superconformal three-point functions
with one chiral primary is necessary to make logical sense
of equation \rr{ProjectorEq}.

The property of scalar semishort operators, that they form a module over the chiral ring, has a remarkable consequence
for the spectrum of the theory: It implies the existence of scalar semishorts at \rwa{low} $J$ as well.
Starting with the moment map operator $(X\bar{X})\uu{3\over 4} = \phi\phb$, we act $J$ times with $\phi$ to obtain
a scalar semishort $\phb\phi\uu {J+1}$.  Algebraically, this state could in principle vanish: \it A priori \rm the representation
of the chiral ring on the module of semishorts need not be faithful.  However we have seen already that the scalar
semishort $\phb\phi\uu {J+1}$ is nonvanishing for sufficiently large $J$, using the effective description!  By associativity,
then, it is impossible for \rwa{any} of the intermediate products $\phb\phi\uu {k+1}$ to vanish, for any nonnegative value of $k$.

The presence of scalar semishorts for all $k$ can of course in principle be seen \it via \rm the superconformal index; and in the Appendix we expand the index to several orders and verify the prediction.  However it should be emphasized that the large-$J$ picture
yields the same conclusion with a far less laborious method.

\section{Conclusions}

In this note we have computed the dimensions of certain operators in the ${\cal N} = 2$ superconformal $XYZ$ model in three dimensions,
to first nontrivial order in an expansion in large $R$-charge $J\ll R$ and large $X$-charge $J\ll X\sim {3\over 2}\cc J\ll R$.  To do this, we treated the theory in radial quantization, and used the effective theory on the moduli space of the $X$ branch.  In this theory, both quantum corrections and
higher-derivative interaction terms are suppressed by powers of $|\phi| = |X|\uu{3\over 4}$ when $|\phi|$ is large, and $|\phi|$ scales as $J\uu{+\hh}$.  We have seen that the state with one $\phb$ excitation in the $s$-wave is protected because its superconformal
representation is semi-short \cite{Dobrev:1985qv,Dobrev:1985vh,Minwalla:1997ka,Kinney:2005ej} and has
to candidate state with the correct quantum numbers to play the role of its $\Qbar\sqd$ descendant.  The third-lowest scalar
primary can be understood as two $\phb$ quanta in the $s$-wave, on top of a sea of $(2J\ll R + 2)$ $\phi$-quanta in the $s$-wave.
This state is in a long multiplet and has an energy shift that is proportional to the coefficient of
the first interaction term -- the supersymmetric version of the 
FTPR term.  By
arguments based on unitarity and causality, the coefficient of the super-FTPR term in the action must be positive, 
and as a result, the energy shift must be \rwa{negative}-definite.  There is an interesting formal similarity between the
large-$R$-charge expansion of the anomalous dimension, and the large-spin expansion of the anomalous dimension
of operators with large spin \cite{Fitzpatrick:2012yx,Komargodski:2012ek}, despite the two expansions resting on rather different logical arguments.
It would be interesting to understand these two expansions within a unified framework of
operator dimensions with large quantum numbers.

One major virtue of the large-$J$ expansion, in the
case where a moduli space exists, is that it gives
us the tools to connect properties of a superconformal field theory that are expressed in "bootstrap"-like language\footnote{For a review of modern developments in the conformal bootstrap, see for instance \cite{Rychkov:2016iqz,Simmons-Duffin:2016gjk} and references therein.} -- \it i.e., \rm anomalous dimensions and operator algebra
structure constants -- with those that can be
expressed in the language of effective field theory
on the moduli space of vacua.  

In moduli space dynamics, superconformal
invariance is spontaneously broken and properties
of the theory can be computed perturbatively in
the low-energy effective theory.  Such perturbative
computations do \rwa{not} rely on any weak coupling
in the underlying dynamics; the perturbative parameter
in the context of moduli space EFT is the ratio of
the infrared to the ultraviolet energy scale, which here
is simply an inverse power of the total charge of the state.
Thus our framework can be used to compute a $1/J$ expansion for properties of near-BPS states in 
a controlled fashion.

As a consistency check, we have verified that the appropriate BPS states with large $X$-charge do in fact exist.  For
chiral ring elements this is immediate, and it is slightly
more nontrivial for semi-shorts scalar states.  Interestingly, once we can verify the existence of scalar semishorts at high $X$-charge, the module structure of the semishorts over the chiral ring, immediately implies the existence of scalar
semishorts at low $X$-charge as well.  This prediction
agrees with explicit calculations of the spectrum extracted with some effort from the superconformal index.  Thus it appears the combination of holomorphy with the large-$J$ expansion is more powerful than the sum of its parts.  It may be hoped that this combination of points of view may
be used to gain insights into the dynamics of other interesting superconformal theories as well.

\section*{Acknowledgments}
\addcontentsline{toc}{section}{Acknowledgments}
The authors are deeply grateful to Thomas Dumitrescu, Daniel Jafferis, and Markus Luty for valuable discussions. The work of SH is supported by the World Premier
International Research Center Initiative (WPI Initiative), MEXT, Japan; by the JSPS Program for Advancing Strategic
International Networks to Accelerate the Circulation of Talented
Researchers;
and also
supported in part by JSPS KAKENHI Grant Numbers JP22740153, JP26400242.
SM and MW  acknowledge the support by JSPS Research Fellowship for Young Scientists.
SH is grateful to the Harvard Center for the Fundamental Laws of Nature, the Burke Institute at Caltech, and the Galileo Galilei Institute during the
"Conformal Field Theories and Renormalization Group Flows in Dimensions $d>2$" conference, for hospitality while this work was in progress.
Some of the calculations of this work were done with the help of the excellent  {Mathematica\texttrademark}  
package {\tt diffgeo.m} developed by M. Headrick \cite{diffgeo}.


\appendix
\section{Notation}
\label{notation}
Here we summarize the notation used in the bulk of this paper. 
The  metric on flat $\IR^{1,2}$ is $\eta_{\mu\nu}=\operatorname{diag}\left(-,+,+\right)$
with $\mu=0,1,2$.
The Dirac matrices 
$(\gamma^\mu)_{\alpha}{}^\beta$   satisfy the Clifford algebra,
	\begin{align}
\left\{\gamma^\mu,\gamma^\nu\right\}_\alpha{}^\beta=(\gamma^\mu)_\alpha{}^\delta (\gamma^\nu)_\delta{}^\beta
+(\gamma^\nu)_\alpha{}^\delta (\gamma^\mu)_\delta{}^\beta = 
2\eta^{\mu\nu}\delta^\beta_\alpha.
	\end{align}
Then, $\gamma^\mu_{\alpha\beta}:=(\gamma^\mu)_\alpha{}^\delta \epsilon_{\delta\beta}$
is 
symmetric in $\alpha\leftrightarrow \beta$.
One may choose $(\gamma^\mu)_{\alpha}{}^\beta = \left(i\sigma^2,\sigma^1,\sigma^3\right)$, so that $\left(  \gamma^\mu \right)^*=\gamma^\mu$.
We define the complex conjugation on products of Grassmann variables as
$\left(\psi_1\psi_2\right)^*=\bar{\psi}_1\bar{\psi}_2$.
A chiral superfield $\Phi\left(  x,\theta,\bar\theta \right)$ is defined by $\bar{\mathcal D}_\alpha \Phi = 0$,
where ${\mathcal D}_\alpha$ and $\bar{{\mathcal D}}_\alpha$ are the  superderivatives,
\begin{align}
{\mathcal D}_\alpha :=\frac{\partial}{\partial\theta^\alpha}
-\left(\gamma^\mu\right)_\alpha{}^\beta\bar\theta_\beta\frac{\partial}{\partial x^\mu}
,\qquad
\bar{\mathcal D}_\alpha :=\frac{\partial}{\partial\bar\theta^\alpha}
-\left(\gamma^\mu\right)_\alpha{}^\beta\theta_\beta\frac{\partial}{\partial x^\mu}.
\end{align}
They satisfy the anticommutation relation,
\begin{align}
	\begin{split}\label{DDAC}
\left\{ {\mathcal D}_\alpha,\bar{\mathcal D}_\beta  \right\}=
\left\{ {\mathcal D}_\beta,\bar{\mathcal D}_\alpha  \right\}=
2\left(  \gamma^\mu \right)_{\alpha\beta}\partial_\mu.
	\end{split}
\end{align}
$\Phi\left(  x,\theta,\bar\theta \right)$ is expanded as
\begin{align}
	\begin{split}\label{ChiralSFExpansion}
\Phi\left(  x,\theta,\bar\theta \right)
={}&\phi(x)+\sqrt2 \theta\phi (x)+\theta^2 F(x)
\\&-\left(   \theta\gamma^\mu\bar\theta\right)\partial_\mu\phi(x)-\frac{1}{\sqrt2}\theta^2\left(\bar\theta\gamma^\mu\partial_\mu\psi(x)\right)+\frac14 \theta^2\bar\theta^2\partial_\mu\partial^\mu\phi(x).
	\end{split}
\end{align}
The normalization for the Berezinian integral is
\begin{align}
\int  \theta^2\bar\theta^2d^2\theta d^2\bar\theta = 1,
\end{align}
and it is convenient to note that, up to total derivatives,
\begin{align}
\int{\mathcal I} d^2\theta d^2\bar\theta  =\left. \frac{1}{16}{\mathcal D}^2\bar{\mathcal D}^2{\mathcal I}\right|_{\theta=\bar\theta=0}.
\end{align}

\section{Uniqueness of the super-FTPR operator on flat space}

We would like to show here that on flat space  there is no supersymmetric dimension-$3$ operator 
 constructed with 
four superderivatives,
 except for the super-FTPR operator \rr{SuperFTPRFlatSpace}.
First of all, we do not have to consider operators containing 
any odd number of superderivatives acting on a single
$\Phi$ or $\bar\Phi$,
because such operators are always equal to ones containing only
even number of superderivatives acting on a single
$\Phi$ or $\bar\Phi$, modulo 
the
leading-order
superspace
equations of motion, ${\mathcal D}^2\Phi \simeq 0$ and 
$\bar{\mathcal D}^2\bar\Phi \simeq 0$.

From (\ref{DDAC}), we have
\begin{align}
	\begin{split}
\bar{\mathcal D}_\alpha{\mathcal D}_\beta \Phi = 
\bar{\mathcal D}_\beta{\mathcal D}_\alpha \Phi,
	\end{split}
\end{align}
and especially $\bar{\mathcal D}^\alpha {\mathcal D}_\alpha \Phi = 0$.
These identities are useful in decreasing the number of index structures.
For instance, one can show that any candidate containing four superderivatives acting on a single $\Phi$
in any order vanishes modulo  the
leading-order
superspace
equations of motion.

By the above consideration, we conclude that only the following operators possibly survive,
\begin{align}
	\begin{split}
{\mathcal  O}_{1}^{(4)}&:=\int d^2\theta d^2\bar\theta
\left(
 \frac{\bar{\mathcal  D}_\alpha {\mathcal  D}_\beta \Phi \bar{\mathcal  D}^\alpha {\mathcal  D}^\beta \Phi}{ \Phi ^3  \bar\Phi } + \textrm{c.c.} \right)  ,
\\
{\mathcal  O}_{2}^{(4)}&:=\int d^2\theta d^2\bar\theta
\frac{ \bar{\mathcal  D}_\alpha {\mathcal  D}_\beta \Phi {\mathcal  D}^\alpha \bar{\mathcal  D}^\beta \bar\Phi }{\left(\Phi\bar\Phi\right)^2}. 
	\end{split}
\end{align}
${\mathcal  O}_{2}^{(4)}$ is equivalent
to the super-FTPR operator  \rr{SuperFTPRFlatSpace}, since $\left\{ {\mathcal D}_\alpha, \bar{\mathcal D}_\beta\right\} = 2\gamma^\mu_{\alpha\beta}\partial_\mu$.
${\mathcal  O}_{1}^{(4)}$ is also equivalent to  the super-FTPR operator,
because
\begin{align}
 \frac{\bar{\mathcal  D}_\alpha {\mathcal  D}_\beta \Phi \bar{\mathcal  D}^\alpha {\mathcal  D}^\beta \Phi}{ \Phi ^3  \bar\Phi }
 \sim \frac{\partial_\mu\Phi \partial^\mu\Phi}{\Phi^3\bar\Phi}
 \sim \frac{1}{\Phi^2}\partial_\mu\left(\frac{\partial^\mu\Phi}{\bar\Phi}\right)
 \sim \frac{\partial_\mu\Phi\partial^\mu\bar\Phi}{\left(\Phi\bar\Phi\right)^2}.
\end{align}
Here, by "$\sim$" we mean modulo total superderivatives, the leading-order equations
of motion,
and numerical coefficients.
So, there is only one supersymmetric dimension-$3$ operator with four superderivatives
on flat space modulo total superderivatives, and it is nothing but the unique super-Weyl completion of the FTPR operator.

\section{Energy correction to one-boson one-fermion and two-fermion excitations}
\label{2b2f4f}

We assume the radius of $S^2$ to be unity throughout this section.

\subsection{One-boson one-fermion excitation}\label{OneFermOneBos}

Quantization of the lowest spin $\psi^\alpha$ state
must be done as follows:
\begin{align}\begin{split}
&\psi^\alpha(x)=\sum_{s=\pm}\beta_s u^\alpha_s(x)+\gamma^\dagger_s v^\alpha_s(x)\label{qc}\\
&\slashed{\nabla}_{S^2}u_s=iu_s,\quad \slashed{\nabla}_{S^2}v_s=-iv_s\\
&\bar u^{(-)} =v^{(-)},\quad \bar u^{(+)} = -v^{(+)}\\
&\gamma^0 u^{(-)}=v^{(+)},\quad \gamma^0 v^{(+)}=-u^{(-)}
\end{split}
\end{align}
We also make use of the equation of motion
for $\psi$:
\begin{equation}
(\gamma^0)^\alpha_\beta\partial_t\psi^\beta
+(\gamma^i)^\alpha_\beta\nabla_{S^2,i}\psi^\beta=0
\end{equation}
At the end of the quantization process we set
\begin{equation}
\bar{u}_\alpha^{s} u^\alpha_r=\frac{1}{4\pi}\delta^{sr},\quad
\bar{u}_\alpha^{s} v^\alpha_r=0
\end{equation}
as a normalisation condition.
According to this quantization convention we get
the free Dirac Hamiltonian,
\begin{eqnarray}
H_0^{\rm Dirac}=\sum_{s=\pm}\left(\beta_s^\dagger \beta_s
+\gamma^\dagger_s\gamma_s
\right),
\end{eqnarray}
with the commutation relation being
\begin{equation}
\{\beta,\beta^\dagger\}=\{\gamma,\gamma^\dagger\}=1.
\end{equation}
The interaction term in the Hamiltonian of order $J^{-3}$
which includes 2-fermion 2-boson interaction is given by
\begin{equation}
\alpha H^{(2,2)}_{\rm int}
=-4\pi\alpha\times 4\bar\psi^\alpha
\left[\left(-\partial_t^2+\nabla^2_{S^2}-{3i}\partial_t+{2}\right)\frac{\bar{F}}{\bar\phi_0^3}\right]
\left[\left(\gamma^\mu_{\alpha\beta}\nabla_\mu + {i}
\gamma^t_{\alpha\beta} \right)\frac{F}{\phi_0^3}\right]\psi^\beta,
\end{equation}
where $\alpha$ is a proportionality constant as in (\ref{bosonicint}).
Making use of the equation of motion and
the fact that $\phi_0=e^{it/2}\varphi_0$
and then taking only the spin-$1/2$ and spin-0 contribution
for the fermion and the boson field, respectively,
we get
\begin{equation}
H^{(2,2)}_{\rm int}
=-\frac{4\pi}{|\varphi_0|^6}\times 
24\bar{\psi}\gamma^0\psi\times(\bar{f}-i\dot{\bar{f}})f
\end{equation}
Using the quantization of the boson field given in Section \ref{quant}
and that of the fermion field given above,
we get
\begin{equation}
H^{(2,2)}_{\rm int}
=-\frac{6}{\pi|\varphi_0|^6}(2a^\dagger-b)a\times
\sum_{s=\pm}\left(\beta_s^\dagger \beta_s
+\gamma^\dagger_s\gamma_s
\right),
\end{equation}
which leads to
the energy correction to the state
$a^\dagger\beta^\dagger_+\Ket{0}$
is
\begin{equation}
\Delta E
=-\frac{12\alpha}{\pi|\varphi_0|^6}.
\end{equation}
This agrees with the energy
correction to the two-boson state,
as it should be from supersymmetry.

\subsection{Two-fermion excitation}\label{TwoFermEx}

The interaction term in the Hamiltonian of order $J^{-3}$
which includes 4-fermion interaction is given by
\begin{equation}
\alpha H^{\rm (0,4)}_{\rm int}=
4\pi\alpha\times\frac{\bar\psi_\beta\bar\psi^\beta}{\bar\phi^3}
\left(-\partial_t^2+\nabla^2_{S^2}-\frac{1}{4}\right)\frac{\psi^\alpha\psi_\alpha}{\phi^3},
\end{equation}
where $\alpha$ is a proportionality constant as in (\ref{bosonicint}).
Making use of the fact that $\phi_0=e^{it/2}\varphi_0$
and taking only the spin-$1/2$ contribution for the fermion field,
we get
\begin{equation}
H_{\rm int}^{(0,4)}=
\frac{4\pi}{|\varphi_0|^6}
\bar{L}(2L+3i\dot{L}-\ddot{L}),
\end{equation}
where $L=\psi\psi$.
Then by using (\ref{qc}) and the normalization condition, we get
\begin{align}\begin{split}
\bar{L}L&=-\frac{1}{8\pi^2}
(\gamma_-\gamma_+
+\beta^\dagger_-\beta^\dagger_+)
(\gamma_+^\dagger\gamma^\dagger_-
+\beta_+\beta_-)\\
\bar{L}\dot L&=
\frac{i}{4\pi^2}(\gamma_-\gamma_+
+\beta^\dagger_-\beta^\dagger_+)
(\gamma_+^\dagger\gamma^\dagger_-
+\beta_+\beta_-)\\
\bar{L}\ddot L&=\frac{1}{2\pi^2}(\gamma_-\gamma_+
+\beta^\dagger_-\beta^\dagger_+)
(\gamma_+^\dagger\gamma^\dagger_-
+\beta_+\beta_-)
\end{split}\end{align}
and $H_{\rm int}^{(0,4)}$ becomes
\begin{equation}
H_{\rm int}^{(0,4)}=-\frac{12}{\pi|\varphi_0|^6}
(\gamma_-\gamma_+
+\beta^\dagger_-\beta^\dagger_+)
(\gamma_+^\dagger\gamma^\dagger_-
+\beta_+\beta_-)
\end{equation}
and the resulting energy correction to the two-fermion state $\beta_+^\dagger
\beta_-^\dagger\Ket{0}$ is
\begin{equation}
\Delta E=-\frac{12\alpha}{\pi|\varphi_0|^6},
\end{equation}
which agrees with the energy correction to
the two-boson state, as it should be from supersymmetry.

\section{Superconformal index and scalar semi-short multiplets}\label{IndexAndSSS}
In this appendix we check the superconformal index for the $XYZ$ model
to confirm that scalar semi-short multiplets really exist in the theory.
The superconformal index for the $XYZ$ model is given by
the following  plethystic exponential \cite{Imamura:2011su,Krattenthaler:2011da},
\begin{align}\begin{split}\label{SCIndex}
I_{XYZ}(x,t_{X},t_{YZ}):={}&\operatorname{Tr}\left((-1)^F z^{\Delta-R-s^3}x^{\Delta+s^3} t_X^{J_X}t_{YZ}^{J_{YZ}}\right)
\\
={}&\exp\left(\sum_{n=1}^\infty \frac{1}{n} F(x^n,t_X^n,t_{YZ}^n)\right),
\end{split}\end{align}

Here, $s^3$ is the third component of the spin on $S^2$, and 
 $t_X$ and $t_{YZ}$ are the fugacities for $U(1)_X$ and $U(1)_{YZ}$, respectively.
$F(x,t_X,t_{YZ})$ is the so-called letter index,
\begin{align}\begin{split}
F(x,t_X,t_{YZ}):= f(x,t_X)+f\left(x,t_X^{-1/2}t_{YZ}\right)+f\left(x,t_X^{-1/2}t_{YZ}^{-1}\right),
\end{split}\end{align}
\begin{align}\begin{split}
f(x,t):= \frac{t x^{2/3}-t^{-1}x^{4/3}}{1-x^2}.
\end{split}\end{align}
Because of the superconformal algebra, only protected multiplets can contribute
to the superconformal index and therefore it is independent of the variable $z$ in (\ref{SCIndex}).
In a BPS multiplet, the BPS primary operator and its $\puu$-derivatives contribute to the index, whereas contributions from the other states in the BPS multiplet cancel between themselves.
In a given scalar semi-short multiplet,
it is the $\bar Q_\uparrow$ descendant of the semi-short primary operator
and its $\puu$-derivatives 
that contribute to the index.
When the index is expanded with respect to $x$,
contributions from BPS multiplets have positive coefficients,
whereas those from semi-short multiplets have negative coefficients.
The operators $\pp\ll{\uparrow\uparrow}$ and $Q\ll{\uparrow}$ are the partial derivatives and supercharges with spin $+1$ and $+\hh$ along
the $z$-axis.

In principle, the superconformal index contains all information about operators in short or semi-short representations.  In principle, 
even the terms in a full and explicit calculation of the index do not necessarily correspond one-to-one with operators satisfying \rr{BPSOp}, because cancellations can occur.  Chiral
ring elements contribute with a positive sign, while
superpartners of scalar semishorts, for instance, contribute with
a negative sign.  In practice, cancellations occur frequently in many
familiar theories, including the $XYZ$ model in $D=3$.  These
cancellations can be removed by organizing the index into
characters of the particular short representations that appear.  

We
do not do this, since the organization into characters is cumbersome a and we are just calculating the some particular terms in the index to establish its agreement with the spectrum of semishort representations as computed with the large-$J$ effective theory.  Rather, we list both the positive and negative contributions 
to the index separately, noting the cancellations as they occur.

To see the existence of the scalar semi-short multiplets,
we expand the superconformal index (\ref{SCIndex}) with respect to $x$
up to and including $O(x^{10/3})$.
However, some contributions from semi-short multiplets are canceled by those
from BPS multiplets, and therefore we cannot see 
all the contributions from semi-short multiplets just by  expanding the index.
So, we  separate these two kinds of contributions order by order,
by brute force.
We also identify all the positive contributions
up to and including $O(x^{10/3})$
 with (descendants of) BPS operators.
The superconformal index (\ref{SCIndex}) is expanded 
with respect to $x$ as follows:
\begin{align}
	\begin{split}
&I_{XYZ}(x,t_X,t_{YZ})
\\&=\uset{1\vphantom{\frac{1}{t_X^{1/2}   t_{YZ}}}}{\mathbbm{1}}
+x^{2/3} \biggl(
\uset{t_X\vphantom{\frac{1}{t_X^{1/2}   t_{YZ}}}}{X}+
\uset{\frac{t_{YZ}}{t_X^{1/2}}}{Y}
+
\uset{\frac{1}{t_X^{1/2}   t_{YZ}}}{Z}
   \biggr)			
 +x^{4/3} \biggl(
 \uset{ t_X^2\vphantom{ \frac{1}{t_X t_{YZ}^2}}}{X^2}+
 \uset{\frac{t_{YZ}^2}{t_X}\vphantom{ \frac{1}{t_X t_{YZ}^2}}}{Y^2}+
\uset{ \frac{1}{t_X t_{YZ}^2}}{Z^2}
 \biggr)			
  \\& +x^2 \biggl(
  \uset{t_X^3\vphantom{  \frac{t_{YZ}^3}{t_X^{3/2}}}	}{X^3}+
  \uset{\frac{t_{YZ}^3}{t_X^{3/2}}}{Y^3}+
  \uset{\frac{1}{t_X^{3/2} t_{YZ}^3}}{Z^3}\biggr)
  	\uset{-2x^2
	\vphantom{   \frac{	t_{YZ}	}{t_X^{1/2}}}}
	{\bar{Q}_\uparrow{D}_X\\\bar{Q}_\uparrow{D}_{YZ}}
     \\& +x^{8/3} \biggl(
     \uset{t_X^4	\vphantom{  \frac{	t_{YZ}	}{t_X^{1/2}}}	}{X^4}+
    \uset{ \frac{t_{YZ}^4}{t_X^2}\vphantom{  \frac{	t_{YZ}	}{t_X^{1/2}}}}{Y^4}+
   \uset{  \frac{1}{t_X^2t_{YZ}^4}\vphantom{  \frac{	t_{YZ}	}{t_X^{1/2}}}}{Z^4}
     +
     \uset{t_X	\vphantom{  \frac{	t_{YZ}	}{t_X^{1/2}}	}	}{\puu X}
     +
    \uset{ \frac{	t_{YZ}	}{t_X^{1/2}}}{\puu Y}+
    \uset{ \frac{1}{t_X^{1/2}t_{YZ}}\vphantom{  \frac{	t_{YZ}	}{t_X^{1/2}}}}{\puu Z}\biggr)
   -x^{8/3}\biggl(
  { t_X}+
   \frac{t_{YZ}}{t_X^{1/2}}+
   \frac{1}{t_X^{1/2}t_{YZ}}\biggr)		
     \\& 
     +x^{10/3} \biggl(
          \uset{t_X^{1/2}t_{YZ}\vphantom{\frac{t_{YZ}^5}{t_X^{5/2}}}}{X\puu Y}+
     \uset{\frac{1}{t_X}\vphantom{\frac{t_{YZ}^5}{t_X^{5/2}}}}{Y\puu Z}+
     \uset{\frac{t_X^{1/2}}{t_{YZ}}\vphantom{\frac{t_{YZ}^5}{t_X^{5/2}}}}{Z\puu X}+
         \uset{ t_X^5\vphantom{\frac{t_{YZ}^5}{t_X^{5/2}}}}{X^5}+
     \uset{\frac{t_{YZ}^5}{t_X^{5/2}}}{Y^5}+
     \uset{\frac{1}{t_X^{5/2}t_{YZ}^5}\vphantom{\frac{t_{YZ}^5}{t_X^{5/2}}}}{Z^5}+
    \uset{t_X^2\vphantom{\frac{t_{YZ}^5}{t_X^{5/2}}}}{\puu X^2}+
    \uset{ \frac{t_{YZ}^2}{t_X}\vphantom{\frac{t_{YZ}^5}{t_X^{5/2}}}}{\puu Y^2}+
     \uset{\frac{1}{t_X t_{YZ}^2}\vphantom{\frac{t_{YZ}^5}{t_X^{5/2}}}}{\puu Z^2}
   \biggr)
 \\&  -x^{10/3}	\biggl(
 t_X^2+
 \frac{t_{YZ}^2}{t_X}+
 \frac{1}{t_X t_{YZ}^2}
  \biggr)+O\left(x^4\right).		
	\end{split}
\end{align}
The negative contribution at $O(x^2)$ is due to
the $\bar{Q}_\uparrow$ descendants of the moment map operators
${D}_X$ and ${D}_{YZ}$,
which  trivially exist because of the
 $U(1)_X$ and $U(1)_{YZ}$ symmetries.
The negative contributions at $O(x^{8/3})$ and at $O(x^{10/3})$ are
nontrivial, however.
These cannot be descendants of the moment map operators on dimensional
grounds.
For instance, the $-x^{8/3} t_X$ and $-x^{10/3} t_X^2$ terms are naturally identified with the 
which are 
the $\bar{Q}_\uparrow$ descendants of semishort
operators of spin $0$ and dimension $5/3$ and $7/3$, respectively.  In terms of the almost-free $\phi$ variables, 
these semishorts can be represented as $\phi\ll 0 \phb\ll 0 \kket {X}$ and $\phi\ll 0 \phb\ll 0 \kket {X\sqd}$, where as explained in section \ref{VevMeaning}, the state $\kket{X\uu J}$ can be thought of as 
$\phi\ll 0 \uu{{{4J}\over 3}}\kket {0}$.

Heuristically, these semi-short operators can be
thought of as $D\ll X \cdot X$ and $D\ll X\cdot X\sqd$, respectively, where $D\ll X$ is the weight-1 scalar semishort "moment map" operator,
whose descendant is the spin-1 $X$-number current.  However this description is not fully precise, because the leading term in the
OPE of $D\ll X$ with $X\uu J$ is not the semishort operator $D\ll X X\uu J$ but rather the chiral primary $X\uu J$, and the coefficient function
is singular, $|\s|\uu{-1}$.  

We emphasize that, for purposes of understanding the spectrum directly at sufficiently large $J$, the power of supersymmetric representation theory is useful mainly as a convenience: The explicit
computations of the large-$J$ effective theory simply agree with
those of the index, with the spectrum computation becoming more
reliable at large $J$.  

The most important thing we learn
directly is that there is a nonzero scalar semishort in the OPE
of $D\ll X$ with $X\uu J$, for sufficiently large $J$:
\begin{align}
	\begin{split}
D\ll X(\s) X\uu J(0) \ni C\uu{D\ll X \cc X\uu J}{}\ll{X\uu J} |\s|\uu 0 
\cc X\uu J(0) + \cdots\ ,
	\end{split}
\end{align}
where the structure constant $C\uu{D\ll X \cc X\uu J}{}\ll{X\uu J}$
and can be calculated semiclassically from an expectation
value of $D\ll X$ in the state $\kket {X\uu J}$,
and is nonzero.  This gives
information about the index only asymptotically.

Combined with the power of associativity, the existence of semishorts at large $J$ has more consequences: Since the product $\cdot$
defines an associative multiplication, and since $D\ll X \cdot X\uu J
\neq 0$, then all the lower-dimension 
products $D\ll X \cdot X$, $D\ll X \cdot X\sqd$, $\cdots$ must automatically
be nonzero as well.  This is in agreement with the index as
expanded above.  So we see that large-$J$ methods 
combined with associativity, yield information about semishort
operators at low $J$ as well.

\section{Semishort superalgebra}\label{SemishortSuperalgebra}

Let us try to set up a formalism of truncating the superalgebra to a finite degrees of freedom, which are creation and annihilation operators.
This section is useful in understanding the vanishing of the 1-loop energy correction to the BPS and semi-short state with given charge $J$.
For the consistency with the notation in \ref{notation}, upper the indices of the operators with dagger assigned.

\subsection{Commutation relations}

We work in radial quantization -- then we have, as a basic building block of the algebra, $\Delta$, the operator dimension, and $\dag$, the Hermitian conjugation.
We are here doomed to dismiss either $P$ or $K$ at the very least, because of the fact that the bosonic conformal algebra can only be unitary represented in infinite-dimension Hilbert space.
At any rate, however, we are trying to find a truncation of the algebra that contains as many supercharges as possible, under such a constraint.

In order to specify such a subalgebra of the full superalgebra that has the above property, let us fix some conventions.
Denote by $Q\upp{\s\ll \D \s\ll{\rm R}}\ll\a$ the
generator that changes the dimension by ${{\s\ll \D}/ 2}$ units,
and the $R$ charge by $\s\ll{\rm R}$ units.
For instance, the we denote by $Q\upp{++}\ll{\uparrow}$ the generator that raises both dimension $\D$ and $R$-charge
$J\equiv J\ll R$ and also the $3$-component $s\ll 3$
of the angular momentum $s\ll a$.

We would now like to restrict our attention to generators that are preserved by the BPS states, that is, we are only going to consider $Q^{(++)}_\alpha$ and $Q^{(--)}_\alpha$.
Now these two have to be related by conjugation, but as we choose the spinor index convention as
\begin{equation}
\left[s\uu a, Q\upp{\pm\pm}\ll\a\right] = \hh \cc \s\uu a\ll{\b\a} \cc Q\upp{\pm\pm}\ll\b\ .
\end{equation}
Then we have
\begin{equation}
\left(Q\upp{\pm\pm}\ll\a\right)\dag = \pm \cc \e\ll{\a\b} \cc Q\uu{\mp\mp}\ll\b\ .
\end{equation}
Hereafter, by using these conventions, we simplify our notation the following way,
\begin{equation}
\bbq\ll\a \equiv Q\upp{++}\ll\a\ , \quad  \bbq\dag\ll\a = 
\left(Q\upp{++}\ll\a\right)\dag
= \e\ll{\a\b}\cc Q\upp{--}\ll\b\ .
\end{equation}

Now, according to the $\mathcal{N}=2$ superconformal algebra,
we have the following commutation relations for these generators chosen to preserve:
\begin{align}
&\left\{ \bbq\ll\a , \bbq\dag\ll\b\right\} = \d\ll{\a\b}\left(\D - J\right) + \s\uu a\ll{\b\a} \cc s\uu a,
\label{QQdagComm}
&
\\
&\left\{ \bbq\ll\a, \bbq\ll\b\right\} = \left\{\bbq\dag\ll\a, \bbq\dag\ll\b\right\} = 0 ,&\\
&\left[J, \bbq\ll\a \right] = + \bbq\ll\a\ , \quad  [J, \bbq\dag\ll\a ] = - \bbq\dag\ll\a ,&
\\
&\left[\D, \bbq\ll\a \right]= + \hh\bbq\ll\a\ , \quad [ \D, \bbq\dag\ll\a ] = - \hh\bbq\dag\ll\a ,&
\\
&\left[s\uu a , \bbq\ll\a\right] = +\hh \cc \s\uu a\ll{\b\a} \cc \bbq\ll\b  , 
\quad 
[s\uu a , \bbq\dag\ll\a] = - \hh \cc \s\uu a\ll{\a\b} \cc \bbq\dag\ll\b ,& \label{QRotTrans}
\\
&\left[s^a,s^b\right]=i\epsilon^{abc}s^c,\quad 
[J,\Delta]=[J,s^a]=[\Delta,s^a]=0.&
\label{bbq}
\end{align}

\subsection{Oscillator realization}

Let us deal with the case where we have a single free multiplet with the 
transformation law of the $s$-wave mode of a free
antichiral superfield $\phb$ on the $S\uu 2$.
We shall
call the bosonic oscillator $a,a\dag$ and
the fermionic oscillator $\bdag\ll\a$.
Note that here we will take the convention that assigns $b\dag_\alpha$ the same transformation law
under spatial rotations, which is given in \eqref{QRotTrans}. 
Then the oscillator realization of the superalgebra is given by
\begin{align}\begin{split}\label{GeneOsc}
J &= \hh b\dag\ll\a b\ll\a - \hh a\dag a\ 
\ ,
 \\
s\uu a &= \hh \s\uu a\ll{\a\b} \cc\cc 
b\dag\ll\a b\ll\b \ ,
\\
\D &= \hh \cc \adag a + \bdag\ll\a b\ll\a\ .
\\
\bbq\ll\a &= \bdag\ll\a \cc a\ , \qquad \bbq\dag\ll\a = \adag\cc b\ll \a\ . 
\end{split}
\end{align}
where we have taken the oscillators to satisfy
canonical commutation and anticommutation
relations
\bbb
\left[a, a\dag\right] = 1\ , \quad\left \{b\ll\a , b\dag\ll\b\right\} = \d\ll{\a\b}\ .
\eee
The generators 
\eqref{GeneOsc}
satisfy the algebra \eqref{QQdagComm}-\eqref{bbq}. 
On the oscillators they act as
\begin{align}
 [s\uu a , \bdag\ll\a ] &= \hh \cc \s\uu a\ll{\b\a} \cc \bdag\ll\a ,
&
 [ s\uu a , b\ll\a ] &= - \hh \cc \s\uu a\ll{\a\b} \cc b\ll\b .
\\
[J, \adag] &= - \hh \adag\ , & [J,a] &= +\hh a\ ,
\\
[J, \bdag\ll\a ] &= + \hh \bdag\ll\a\ , &  [J, b\ll\a] &= - \hh b\ll\a\ ,
\\
[s\uu a , \bdag\ll\a ] &= \hh \cc \s\uu a\ll{\b\a} \cc \bdag\ll\b ,
& [ s\uu a , b\ll\a ] &= - \hh \cc \s\uu a\ll{\a\b} \cc b\ll\b .
\end{align}

\subsection{Perturbation theory}

Now let us set us perturbation theory.
We wish to make a small
change to the generators, at order $\e$, and demand that the structure
of the superalgebra still be preserved at order $\e$.  So define
$\bbq\ll\a(\e)$, $\bbq\dag\ll\a(\e)$, $\D(\e)$ to be
\begin{align}
	\begin{split}
\bbq\ll\a(\e) &= \bbq\ll\a(0) + \e \bbq\ll\a\pr(0) + O\left(\epsilon^2\right)
,\\
\bbq\ll\a\dag(\e) &= \bbq\ll\a\dag(0) + \e \bbq\ll\a\dag{}\pr(0) + O\left(\epsilon^2\right),
\\
\D(\e) &= \D(0) + \e \D\pr(0) + O\left(\epsilon^2\right).
	\end{split}
\end{align}
Many first-order perturbations of the algebra are unphysical, and correspond
merely to first-order redefinitions of the variables induced by transformations on Hilbert space;
we fix much of this ambiguity by the condition that $s\uu a$ and $J$ to be \rwa{constant}, independent of $\e$:
\begin{align}
	\begin{split}
{{dJ}\over{d\e}} = {{d s\uu a}\over{d\e}} = 0\ .
	\end{split}
\end{align}
Then use the notation
\begin{align}
\bbq\ll\a &\equiv \bbq\ll\a(0) \ ,
 &\bbq\ll\a\dag &\equiv \bbq\ll\a\dag(0),
\\
\qqq\ll\a &\equiv \bbq\ll\a\pr(0) \ ,
&\qqq\dag\ll\a &\equiv \bbq\dag\ll\a{}\pr(0),
\\
\D\pr &= \ddp\pr \equiv \D\pr(0) = \ddp\pr(0),&&
\end{align}
and take the first $\e$-derivative of the the algebra \eqref{QQdagComm}-\eqref{bbq}
\begin{align}
&
\{ \bbq\ll\a(\e) , \bbq\dag\ll\b (\e)\} = \d\ll{\a\b}\ddp(\e) + \s\uu a\ll{\b\a} \cc s\uu a
,&&
\label{QQCommBoring}
\\
&
\{ \bbq\ll\a(\e), \bbq\ll\b(\e)\} = \{\bbq\dag\ll\a(\e), \bbq\dag\ll\b(\e)\} = 0\ ,&&&
\label{QQCommInteresting}
\\
&
[ J, \bbq\ll\a(\e) ] = + \bbq\ll\a (\e) \ , 
&& [J, \bbq\dag\ll\a(\e) ] = - \bbq\dag\ll\a (\e)\ ,&
\\
&
[\D(\e), \bbq\ll\a(\e) ] =  + \hh\bbq\ll\a(\e)\ , 
&& [ \D(\e), \bbq\dag\ll\a(\e) ] = - \hh\bbq\dag\ll\a(\e)\ .&
\label{DeltaQComm}
\\
&
[s\uu a , \bbq\ll\a(\e)] = +\hh \cc \s\uu a\ll{\b\a} \cc \bbq\ll\b(\e) \ , &&
[s\uu a , \bbq\dag\ll\a(\e)] = - \hh \cc \s\uu a\ll{\a\b} \cc \bbq\dag\ll\b(\e)\ ,&
\\
&[ s\uu a , s\uu b ] = + i \cc \e\uu{abc} \cc s\uu c\ ,&&
 [\D(\e), s\uu a] = 0\ ,&
\end{align}
and evaluate at $\e = 0$.
This gives a set of "easy" perturbation equations,
which involve commutators with the fixed generators
$J$ and $s\uu a$,
\begin{align}\label{EasyPertEqs}
[ J, \qqq\ll\a ] &= + \qqq\ll\a \ , 
& [J, \qqq\dag\ll\a ] &= - \qqq\dag\ll\a \ ,
\\
[s\uu a , \qqq\ll\a] &= +\hh \cc \s\uu a\ll{\b\a} \cc \qqq\ll\b \ , 
&
[s\uu a , \qqq\dag\ll\a] &= - \hh \cc \s\uu a\ll{\a\b} \cc \qqq\dag\ll\b\ ,
\\
[s\uu a, \D\pr] &= [s\uu a, \ddp\pr] = 0\ .&&
\end{align}
and "hard" perturbation equations which involve
two different perturbations:
\begin{align}
&\{ \bbq\ll\a, \qqq\ll\b\} + (\a\leftrightarrow\b)  =\{ \bbq\dag\ll\a, \qqq\dag\ll\b\}  + (\a\leftrightarrow\b) = 0\ ,
\label{HardPertEqA}
\\
&\{ \bbq\ll\a , \qqq\dag\ll\b \} + 
\{ \bbq\dag\ll\b ,  \qqq\ll\a  \}   = \d\ll{\a\b}\ddp\pr = \d\ll{\a\b}\Delta\pr \ ,
\label{HardPertEqB}
\\
&
[\D, \qqq\ll\a] - [\bbq\ll\a, \D\pr] 
= + \hh\qqq\ll\a\ ,
\label{HardPertEqC}
\\
&
 [\D, \qqq\dag\ll\a] - [\bbq\dag\ll\a, \D\pr] = - \hh\qqq\dag\ll\a\ .
 \label{HardPertEqD}
\end{align}
The "easy" perturbation equations \rr{EasyPertEqs} just express that
the transformation laws of the perturbed generators under
the "fixed" generators $J$, $s\uu a$ are the same as those of
the corresponding unperturbed generators.

Let us   solve the hard perturbation equations.
First, start with equation \eqref{HardPertEqA}.
Solving this in full generality may be difficult, but can be at least done in a sufficient condition way,
i.e. this equation is solved by
\bbb
\qqq\ll\a \equiv [\bbq\ll\a, \co\ls 2] \ , \llsk\llsk \qqq\ll\a\dag = - [\bbq\dag, \co\dag\ls 2]
\eee
We can use the notation $\cdot$ for acting by commutation or anticommutation.
We denote this by $\rbbq$, and also define
\bbb
\rbbq\sqd \equiv \e\ll{\a\b} \rbbq\ll\a \rbbq\ll\b\ , \llsk\llsk \rbbq\dag{}\sqd \equiv \e\ll{\a\b} \rbbq\dag\ll\a \rbbq\dag\ll\b
\eee
The meaning of the subscript ${}\ls 2$ will become clear shortly.
So then we have 
\begin{align}
	\begin{split}
\qqq\ll\a 
\overset{\purple{\downarrow}}{=} 
\rbbq\ll\a \cdot\co\ls 2 \ , \qquad
\qqq\ll\a\dag = - \rbbq\dag\cdot\co\dag\ls 2\ .
	\end{split}
\end{align}
The symbol $\overset{\purple{\downarrow}}{=} $ means it's just an ansatz.
But this ansatz does automatically solve \rr{HardPertEqA}.  The easy
equations \rr{EasyPertEqs} just constrains $\co$ to be
a scalar with vanishing $R$-charge.

Now consider equation \rr{HardPertEqB}.  Contracting it with $\hh \d\ll{\a\b}$, this equation
tells us that 

\bbb
\D\pr = \hh \rbbq\dag\ll\a \rbbq\ll\a \co\ls 2 - \hh \rbbq\ll\a \rbbq\dag \ll\a \co\dag\ls 2\ .
\een{ForcedAnsatz}
An imaginary part of $\co$ contributes a total derivative to $\D\pr$.  For
real ${\cal A}$,
\bbb
\co\ls 2 \to \co\ls 2 + i \cc {\cal A}\ , \llsk\llsk \D\pr \to \D\pr +  \dot{{\cal A}}\ ,
\eee
so an imaginary part of $\co$ just corresponds to changing the
Hamiltonian by conjugation by an infinitesimal unitary transformation
parametrized by ${\cal A}$ which is scalar and $J$-neutral.
Since ultimately we only care about the system up to change of basis,
we can fix that ambiguity by simply taking the convention
\begin{align}
	\begin{split}
	 \co\dag\ls 2 = \co\ls 2.
\label{RealityConvention}
	\end{split}
\end{align}
 With convention \rr{RealityConvention} we get
\begin{align}
	\begin{split}
\D\pr = \hh [ \rbbq\dag\ll\a,  \rbbq\ll\a ]\cc  \co\ls 2 .
\label{DefiningEqDeltaPr}
	\end{split}
\end{align}
So now equation \rr{HardPertEqB} reads:
\begin{align}
	\begin{split}
[ \rbbq\dag\ll\b , \rbbq\ll\a ] \cdot \co\ls 2 = \d\ll{\a\b} \cc \D\pr\ .
\label{OneHardPertEqn}
	\end{split}
\end{align}
Since we can take equation \rr{DefiningEqDeltaPr} to \rwa{define} $\D\pr$, 
the only remaining content of \rr{OneHardPertEqn} is equivalent to the
statement that
\begin{align}
	\begin{split}
\s\uu a\ll{\a\b} [ \rbbq\dag\ll\b , \rbbq\ll\a ] \cc  \co\ls 2 = 0\ .
	\end{split}
\end{align}\label{EqnToSolve}
The ansatz we're going to make, to solve \eqref{EqnToSolve}, is
\begin{align}
	\begin{split}
\co\ls 2 \overset{\purple{\downarrow}}{=} \hh [\rbbq\ll \g\dag, \rbbq\ll\g] \cdot \co\ls 0
- \cc \co\ls 0 .
\label{SSLikeAnsatz} 
	\end{split}
\end{align}
The first term would be present in flat-space SUSY.  Indeed,
the formula for $\D\pr$ in terms of four supercharges
acting on $\co\ls 0$, is just an operator realization
of superspace perturbation theory, with $\co\ls 0$ playing
the role of the superspace integrand of $D$-term
type.  The second
term on the RHS of \rr{SSLikeAnsatz} is not present
in flat-space SUSY, and corresponds to a nontrivial
background curvature of superspace in the sense of \cite{Dumitrescu:2012ha,Festuccia:2011ws}.

\subsection{A last bit of closure of the algebra}

There is one last nontrivial equation that must be satisfied.  It comes
from eq. \rr{HardPertEqC} (and its conjugate \rr{HardPertEqD}).
Equation \rr{HardPertEqC} reads
\bbb
[\D, \qqq\ll\a] - [\bbq\ll\a, \D\pr] 
= + \hh\qqq\ll\a
\een{HardPertEqCRECAP}
This equation does not impose any further independent conditions on the perturbation of the generators.  In principle
it follows automatically and can be verified directly on the generators constructed from ${\cal O}\ls 0$.  To see this it is
simplest to note that this equation is the commutator \rr{DeltaQComm} at first order in $\e$;
this equation is forced by \rr{QQCommBoring} and \rr{QQCommInteresting} via the graded Jacobi identity, and
this must hold order by order in $\e$.  Thus \rr{HardPertEqCRECAP} follows automatically from the other
first-order closure equations \rr{HardPertEqA} and \rr{HardPertEqB} without imposing further conditions on the perturbation.

\subsection{Examples}

Now we would like to apply our formula to some examples
of $\co\ls 0$, which correspond to the superspace
integrand of superspace perturbation theory, to
see concretely how interaction terms 
made from the semishort zero mode correspond to perturbations $\D\pr$ of the Hamiltonian.  In particular, we will see that
all such perturbations come out automatically normal-ordered, with at least one semishort zero mode
annihilation operator on the right.

\heading{Example: Perturbation corresponding to quadratic deformations}

So the simplest sort of deformation to add would of course be $\co\ls 0 \equiv
\cce\cc
 a\dag a$.
Then using \rr{GeneOsc}, 
we find that 
\begin{eqnarray}
\co\ls 2 = \rbbq\dag\ll\g \rbbq\ll\g\co\ls 0 -  \rrD\co\ls 0 + \k\co\ls 0
=  \rbbq\dag\ll\g \rbbq\ll\g\co\ls 0 + \k\co\ls 0
=  \cce \left(a\dag a + b\dag\ll\g b\ll\g\right)
\end{eqnarray}
and the perturbation of the supercharges and dilatation operator simply vanish:
\begin{align}
	\begin{split}
\qqq\ll\a = \rbbq\dag\ll\a  = \D\pr = 0.
	\end{split}
\end{align}

\heading{Quartic perturbation}

Now let us work out the formulae for the quartic perturbation.
We define
\begin{align}
	\begin{split}
\co\ls 0 = {{\cce}\over 4}\cc a\ddsq a\sqd\ .
	\end{split}
\end{align}
Then
we have
\begin{align}
	\begin{split}
\co\ls 2 
= \cce\cc \left( {3\over 4} a\ddsqd a\sqd + b\dag\ll\g b\ll\g a\dag a
\right)\ ,
	\end{split}
\end{align}
and
\begin{align}
	\begin{split}
\qqq\ll\a  &=   {\cce\over 2} \cc b\dag\ll\a \cc a\dag\cc a\sqd 
    - \cce b\dag\ll\g  b\dag\ll\a b\ll\g\cc \cc a ,
    \\
    \qqq\dag\ll\a
&= {\cce \over 2} \cc a\ddsqd \cc b\ll\a \cc a 
    - \cce b\dag\ll\g  \cc a\dag\cc b\ll\a b\ll\g\cc .
	\end{split}
\end{align}
The first-order modification $\Delta\pr$
of the operator Hamiltonian is
\begin{align}
	\begin{split}
\D\pr = 
  2\cc \cce  a\dag\cc  b\dag\ll\a \cc b\ll\a \cc a 
+\cce  \cc a\ddsqd \cc  a\sqd
    + \cce b\dag\ll\g  \cc b\dag\ll\a\cc b\ll\a b\ll\g
    .
	\end{split}
\end{align}

\heading{More general perturbations with a single semishort multiplet}
\def\ddx#1{^{\dagger {#1}}}
\def\ddu#1{^{\dagger {#1}}}
The most general perturbation ${\cal O}\ls 0$ you can write down made from the
bosonic oscillator, preserving the R-symmetry, is 
\begin{align}
	\begin{split}
{\cal O}\ls 0 \equiv {{\cce}\over{p\sqd}} a\ddx p a\uu p\ .
	\end{split}
\end{align}
So then
\begin{align}
	\begin{split}
\cce\uu{-1}\cc \co\ls 2 
&=\left ({2\over p} - {1\over{p\sqd}} \right )
\cc a\ddx p a\uu p
 +b\dag\ll\g\cc a\ddx {p-1} a\uu {p-1} b\ll\g ,
 \\
 \cce\uu{-1} \cc q\ll\a &=  \left (1 - { 1\over p}  \right )\cc a\ddx{p-1}
\cc b\dag\ll\a \cc a\uu p
- (p-1) \cc b\dag\ll\g\cc b\dag\ll\a \cc  a\ddx {p-2} a\uu {p-1} b\ll\g \ ,
\\
\cce\uu{-1} \cc q\ll\a\dag &=
 \left (1 - { 1\over p}  \right )\cc a\ddx p a\uu{p-1}
\cc b\ll\a 
- (p-1) \cc b\dag\ll\g \cc  a\ddx {p-1} a\uu {p-2} \cc b\ll\a\cc b\ll\g,
\\
\cce\uu{-1}\cc \D\pr &= 2\left (p -1 \right ) \cc b\dag\ll\a
\cc  a\ddx {p-1} \cc a\uu{p-1}
\cc b\ll\a   +
2\cc \left (1 - { 1\over p}  \right )\cc
 \cc a\ddx p a\uu p\\
 &\phantom{=}
 + (p-1)\sqd \cc b\dag\ll\g \cc b\dag\ll\a \cc  a\ddx {p-2} \cc
a\uu {p-2} \cc b\ll\a\cc b\ll\g.
	\end{split}
\end{align}
For $p=1$, note that $q_\alpha$, $a_\alpha^\dagger$ and $\Delta'$ vanish, as we found earlier.
\heading{The BPS zero mode multiplet}

Now we introduce the BPS zero mode.  Let us call it $z$, but we shall
think of it as corresponding to $\phi\ll 0$, up to a constant.

From the point of view of our small superalgebra, this operator $z$
is actually a rather funny
object.  It has $J = +\hh$ and at the free level, it has $\D = + \hh$ too.
This means it commutes with $\D - J$.  Since it is a BPS primary field, it also
commutes with the $\bbq$ and $\bbq\dag$ as well.  So, from the point
of view of the small superalgebra, $z$ is really just a $c$-number.  However,
on the other hand, $z$ only commutes with $\D - J$, and not with
$\D$ and $J$ individually.  So, since we have not yet specified the
normalization of $z$, let us define it so that
\begin{align}
	\begin{split}
[z\dag, z]= +1\ .
	\end{split}
\end{align}
Note that this is only possible in a unitary theory because $z$ is the
energy-raising, rather than the energy-lowering part of $\phi$.

So, $z$ has the same $R$-charge as $a$, but the same frequency
as $a\dag$.  The composite object $\hat{A}\dag \equiv z\adag$ has frequency 
$1/r$ and vanishing $R$-charge.  We can therefore make
new interesting perturbations out of this operator.

Since $z, z\dag$ commute with the whole superalgebra, $\hat{A}$ and 
$\hat{A}\dag$ have the same SUSY representations as $a, a\dag$ respectively.
Defining $\hat{B}\dag\ll\a \equiv z b\dag\ll\a$, we have
\begin{align}
&\rbbq\ll\a \hat{A} = 0 \ ,&& \rbbq\dag\ll\a \hat{A} = - \hat{B}\ll\a\ ,
\\
&\rbbq\ll\a \hat{A}\dag = \hat{B}\ll\a \ , &&\rbbq\dag\ll\a \hat{A}\dag = 0 \ ,
\\
&\rbbq\ll\a \hat{B}\ll\b = \d\ll{\a\b}\cc \hat{A}\ ,&& \rbbq\ll\a\dag \hat{B}\ll\b = 0\ ,
\\
&\rbbq\ll\a \hat{B}\dag\ll\b = 0 \ ,&& \rbbq\dag\ll\a \hat{B}\dag\ll\b = \d\ll{\a\b}\cc \hat{A}\dag\ .
\end{align}

\heading{General bosonic perturbations involving a semishort and a BPS multiplet}

So now we can make all sorts of fascinating R-symmetric perturbations such
as
\begin{align}
	\begin{split}
{\cal O}\ls 0 = {1\over{pq}}\cc \hat{A}\ddx q \hat{A}\uu p\ .
	\end{split}
\end{align}
This is non-Hermitean, but we can always add the Hermitean conjugate
to make it Hermitean.
So we have
\begin{align}
	\begin{split}
\rbbq\ll\g {\cal O}\ls 0 = {1\over p} \cc \hat{B}\dag\ll\g\cc \hat{A}\ddx{q-1} \cc \hat{A}\uu p
	\end{split}
\end{align}
and
\begin{align}
	\begin{split}
\rbbq\dag\ll\g \rbbq\ll\g {\cal O}\ls 0 = 
+
  {2\over p} \cc  \hat{A}\ddx q \cc \hat{A}\uu p
  + \hat{B}\dag\ll\g\cc \hat{A}\ddx{q-1} \cc \hat{A}\uu {p-1} \hat{B}\ll\g.
	\end{split}
\end{align}

One major difference, now that $z, z\dag$ have been introduced, 
is that $R$-neutral operators no longer necessarily commute with $\D$.
In particular we have
\begin{align}
&[\D, \hat{A}\dag] = +\hat{A}\dag\ , && [\D, \hat{A}] = - \hat{A}\ ,
\\
& [\D, \hat{B}\dag] = +{3\over 2}\hat{B}\dag\ , && [\D, \hat{B}] = - {3\over 2}\cc \hat{B}\ , 
\end{align}
and so
\begin{align}
	\begin{split}
\left[\D, \hat{A}\ddx q \hat{A}\uu p\right] = (q-p) \hat{A}\ddx q \hat{A}\uu p\ .
	\end{split}
\end{align}
The expressions for $\co\ls 2$ and the perturbed generators are:
\begin{align}
	\begin{split}
{\cal O}\ls 2 
&= 
\rbbq\dag\ll\g \rbbq\ll\g {\cal O}\ls 0 - \rrD{\cal O}\ls 0 + \k\co\ls 0
 = \frac{p+q-1}{pq}
\cc  \hat{A}\ddx q \cc \hat{A}\uu p
+ \hat{B}\dag\ll\g\cc \hat{A}\ddx{q-1} \cc \hat{A}\uu {p-1} \hat{B}\ll\g
,
\\
\qqq\ll\a &= \rbbq\ll\a  {\cal O}\ls 2 
= 
\frac{q-1}{p}
\cc   \hat{B}\dag\ll\a \cc \hat{A}\ddx {q-1} \cc \hat{A}\uu p
-(q-1) \hat{B}\dag\ll\g\cc \hat{B}\dag\ll\a
\cc \hat{A}\ddx{q-2} \cc \hat{A}\uu {p-1} \hat{B}\ll\g
,
\\
\qqq\dag\ll\a &= \rbbq\dag\ll\a\co\ls 2
= 
\frac{p-1}{q}
\cc    \hat{A}\ddx q \cc \hat{A}\uu {p-1}\cc\hat{B}\ll\a
-(p-1) \hat{B}\dag\ll\g\cc 
\cc \hat{A}\ddx{q-1} \cc \hat{A}\uu {p-2}\hat{B}\ll\a \hat{B}\ll\g
,	\\
\D\pr &= \left ( p + q -2 \right )
\cc   \hat{B}\dag\ll\a\cc \hat{A}\ddx {q-1} \cc \hat{A}\uu {p-1}\cc\hat{B}\ll\a
+
\left ( {q\over p} + {p\over q}  - {{p+q}\over{pq}} \right )
\cc    \hat{A}\ddx q \cc \hat{A}\uu p\cc
\\&\phantom{=}
+(p-1)(q-1) \hat{B}\dag\ll\g\cc 
\cc\hat{B}\dag\ll\a\cc
\hat{A}\ddx{q-2} \cc \hat{A}\uu {p-2}\hat{B}\ll\a \hat{B}\ll\g
.
	\end{split}
\end{align}
This formula of course assumes neither $p$ nor $q$ vanishes; we normalized
the perturbation, for convenience, by dividing by $pq$ at the beginning.  If 
instead we hadn't, and we had defined
\begin{align}
	\begin{split}
\co\ls 0 \equiv a\ddx q a\uu p\ ,
	\end{split}
\end{align}
then we would have had
\begin{align}
	\begin{split}
\D\pr = {}&
pq \left ( p + q -2 \right )
\cc   \hat{B}\dag\ll\a\cc \hat{A}\ddx {q-1} \cc \hat{A}\uu {p-1}\cc\hat{B}\ll\a
+
\left ( q\sqd + p\sqd  - p-q \right )
\cc    \hat{A}\ddx q \cc \hat{A}\uu p\cc
\\&+(p-1)(q-1) \hat{B}\dag\ll\g\cc 
\cc\hat{B}\dag\ll\a\cc
\hat{A}\ddx{q-2} \cc \hat{A}\uu {p-2}\hat{B}\ll\a \hat{B}\ll\g
.
	\end{split}
\end{align}

\heading{Protection of the semishort state}

So now we see, \rwa{regardless of the form of the perturbation}, the
Hamiltonian perturbation not only has vanishing expectation value in
the semishort multiplet, it simply \rwa{annihilates} the entire
semi-short multiplet, as SM pointed out.  This much stronger condition would
seem to guarantee the protection of the semishort multiplet not just to
first order, but to all orders in perturbation theory.

The nonzero modes of the free antichiral superfield $\phb$ and the free chiral superfield $\phi$
should also be included; from the point of view of quantum mehchanics, these are
higher-spin multiplets, obtained by Kaluza-Klein reduction of the $2+1$-dimensional 
superfields on the sphere. 

\bibliographystyle{utphys-edited} 
\bibliography{references-POST-SUBMISSION}

\end{document}